\documentclass[a4paper,superscriptaddress,twocolumn,aps]{revtex4-1}
\pdfoutput=1

\usepackage{amsmath,,amssymb}
\usepackage{tikz}
\usetikzlibrary{positioning}
\usepackage{graphicx}
\usepackage{dcolumn}
\usepackage{bm}
\usepackage{subfigure}
\usepackage{lipsum}
\usepackage{mathtools}

\usepackage[colorlinks=true,citecolor=blue, urlcolor=blue, linkcolor=blue]{hyperref}

\newcommand{\cbox}[2]{\vcenter{\hbox{\includegraphics[width=#1em]{#2}}}}
\newcommand{\z}[1]{$\mathbb{Z}_#1$}

\newcommand{\ket}[1]{\left\vert#1\right\rangle}
\newcommand{\bra}[1]{\left\langle#1\right\vert}

\usepackage{color}

\begin{document}

\title{Gapped \z{2} spin liquid in the breathing kagome Heisenberg antiferromagnet}

\author{Mohsin Iqbal}
\affiliation{Max-Planck-Institut f{\"u}r Quantenoptik, Hans-Kopfermann-Stra{\ss}e 1, 85748 Garching, Germany}
\affiliation{Munich Center for Quantum Science and Technology, Schellingstra{\ss}e 4, 80799 M{\"u}nchen, Germany}

\author{Didier Poilblanc}
\affiliation{Laboratoire de Physique Th\'eorique, IRSAMC, Universit\'e de Toulouse, CNRS, UPS, France}

\author{
Norbert Schuch}
\affiliation{Max-Planck-Institut f{\"u}r Quantenoptik, Hans-Kopfermann-Stra{\ss}e 1, 85748 Garching, Germany}
\affiliation{Munich Center for Quantum Science and Technology, Schellingstra{\ss}e 4, 80799 M{\"u}nchen, Germany}

\begin{abstract}
	{We investigate the spin-1/2 Heisenberg antiferromagnet on the
kagome lattice with breathing anisotropy (i.e. with weak and strong
triangular units), constructing an improved simplex Resonating Valence
Bond (RVB) ansatz by successive applications (up to three times) of local
quantum gates which implement a filtering operation
	on the bare nearest-neighbor RVB state. The resulting Projected
Entangled Pair State involves a small number of variational parameters
(only one at each level of application) and preserves full lattice and
spin-rotation symmetries. Despite its simple analytic form, the simplex
RVB provides very good variational energies at strong and even intermediate
 breathing anisotropy. We show that it carries \z{2} topological order which does not fade away under the first few applications of the quantum gates, suggesting that the RVB topological spin liquid becomes a competing ground state candidate for the kagome antiferromagnet at large breathing anisotropy. 
	 }
\end{abstract}
\maketitle

\section{Introduction}

The Resonating Valence Bond (RVB) state, defined as an equal weight superposition of (non-orthogonal) nearest neighbor (NN) singlet (or dimer) coverings, was
first proposed by Anderson~\cite{Anderson1973} to describe a possible spin liquid ground
state (GS) of the $S=1/2$ antiferromagnetic Heisenberg (HAF) model on the triangular lattice. Later on, it was also introduced as the parent Mott state of 
high-$T_c$ superconductors~\cite{RVB}.
Several 
works ~\cite{poilblanc2012topological,schuch2012resonating,wildeboer,yangfan} have demonstrated that NN RVB states defined on triangular and kagome lattices
are gapped spin liquid states with $\mathbb{Z}_2$ topological order, and GSs of local parent Hamiltonians~\cite{schuch2012resonating,zhou2014}. 

Spin liquid behaviors are expected in two-dimensional (2D) frustrated quantum magnets where magnetic frustration prohibits magnetic ordering at zero temperature. 
The spin-1/2 Heisenberg antiferromagnet on the kagome lattice (KHAFM) is believed to be the simplest archetypical model hosting a spin liquid GS with no Landau-Ginzburg spontaneous symmetry breaking. However, the precise nature of this spin liquid is still actively debated. While the HLSM theorem \cite{hastings2004lieb} excludes a unique GS separated from the first excitations by a finite gap (so-called ``trivial'' spin liquid), 
a gapless spin liquid  \cite{Ran2007,iqbal2013gapless,he2017signatures,liao2017gapless} or a gapful {\it topological} spin liquid (of the RVB type)  \cite{yan2011spin,mei2017gapped,depenbrock:kagome-heisenberg-dmrg} are the two favored candidates.

\begin{figure}[t!]
	\centering
	\includegraphics[width=16em]{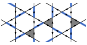}
	\caption{A singlet covering of RVB state on the kagome lattice. Arrowheads indicate the counter-clockwise orientation of singlets on edges. 
	Gray indicates defect triangles of the singlet covering.}
	\label{fig:kagome_1}
\end{figure}

An important aspect is to understand the stability of the spin liquid GS
against various perturbations, such as long-range interactions or
different anisotropies. Beyond being of interest by itself, 
it might also yield alternative ways to assess the nature of the ground
state of the isotropic KHAFM by allowing to adiabatically connect it to a limiting case
which might be easier to study.  An important case of such perturbations
is the HAF model on the kagome lattice with anisotropy, which can be
written as
\begin{equation}\label{eq:heis1}
H(\gamma) =  \sum_{(ij) \in \triangleright }{\textbf{S}_i . \textbf{S}_j}
+  \gamma \sum_{ (ij) \in \triangleleft }{\textbf{S}_i . \textbf{S}_j}
\end{equation}
with $0\le\gamma\le1$ (where $S_z=\pm\tfrac12$). The Hamiltonian $H(\gamma)$, except at $\gamma=1$, explicitly breaks the inversion
symmetry between the strong (or right-pointing) $\triangleright$ and 
the weak (or left-pointing) $\triangleleft$ triangles of the kagome
lattice (Fig.~\ref{fig:kagome_1}). The anisotropic model~(\ref{eq:heis1})
(also referred to as ``breathing'' HAF \cite{okamoto2013breathing}) has
gained additional relevance because recent studies have shown a realization of
\eqref{eq:heis1} for particular values of $\gamma$ in a vanadium-based
compound~\cite{Aidoudi2011,Harrison2013,Bert2017}.  Moreover, in the limit
of strong anisotropy, $\gamma\to0$, it can be mapped to a simpler model
with two spin-$\tfrac12$ degrees of freedom per site, 
similar to a Kugel-Khomskii model~\cite{mila:breathing}.

The Hamiltonian~\eqref{eq:heis1}
has been studied using different numerical methods: In Ref.~\onlinecite{schaffer2017quantum}, Gutzwiller-projected generalized BCS wavefunctions have been used, finding a gapped $\mathbb Z_2$ topological phase throughout; in contrast to this, Ref.~\onlinecite{iqbal2018persistence}, supplementing the same ansatz with two Lanczos steps and anisotropic couplings in an enlarged unit cell, finds that around the isotropic point $\gamma=1$, a gapless $U(1)$ Dirac spin liquid (DSL) phase outperforms the gapped $\mathbb Z_2$ phase for sufficiently large systems, while for 
$\gamma\lesssim 0.25$, Valence Bond Crystal (VBC) order dominates. Finally, Ref.~\onlinecite{repellin2017stability} analyzes the model using iDMRG, supplemented by exact diagonalization, and finds a $\mathrm{U}(1)$ DSL for sufficiently large $\gamma$, which at $\gamma\lesssim 0.1$ transitions to a phase with nematic order (i.e., breaking lattice rotation symmetry).  In the light of these conflicting results, the nature of the strongly anisotropic limit, and the question whether it is adiabatically connected to the isotropic KHAFM seems wide open.

In this paper, we use an ansatz based on a systematic optimal cooling
procedure, applied to the RVB state, to analyze the nature of the
breathing KHAFM, focusing on the strong anisotropy limit. Our ansatz,
which we term ``simplex RVB'', clearly outperforms the previous results
obtained for the thermodynamic limit, and clearly yields a gapped
$\mathbb Z_2$ spin liquid rather than a VBC phase. Our ansatz differs from
previous approaches in several ways: First, it implements a systematic and
optimized cooling procedure -- in essence, an optimized imaginary time
evolution scheme -- which can be systematically constructed from any
Hamiltonian at hand. Second, it requires only a very small number of
parameters with a clear physical interpretation; in our case, we use at
most $3$ parameters. Third, those parameters  have a clear physical
interpretation in terms of a variational RVB-type wavefunction: Their role
is to create longer-range singlets with suitable amplitude and phase such
as to systematically decrease the energy of the variational wavefunction.
And lastly, the clear role of the variational parameters in the ansatz
facilitates the analysis of its order.

Our analysis reveals a gapped $\mathbb Z_2$ topological spin liquid phase
for the whole range $0\le\gamma\le1$.  In particular, in the strongly
anisotropic limit, our results clearly outperform the energies previously
obtained in the thermodynamic limit~\cite{iqbal2018persistence} which
found a VBC phase, while at the same time they require a significantly
smaller number of variational parameters. More specifically, our ansatz
with $2$ parameters -- corresponding to only one optimized trotterized
imaginary time evolution step on top of the RVB state -- already yields a slightly better
energy than the VBC ansatz with $2$ Lanczos steps, while it clearly
outperforms it with an additional parameter (half a Trotter step). This can
be attributed to the fact that our ansatz, unlike Lanczos steps, captures
the extensive nature of perturbations and thus correctly reproduces the
perturbative expansion in the thermodynamic
limit~\cite{vanderstraeten:peps-perturbations}.

On a technical level, we use the formalism of Projected Entangled Pair
States (PEPS)~\cite{verstraete2004renormalization} to implement the
simplex RVB ansatz.  The idea of the PEPS description is to specify the
entanglement structure of the wavefunction as a network of local tensors.
The so-called bond dimension determines the efficiency of the PEPS
description. The NN RVB state on the kagome lattice can be represented as
a PEPS with bond dimension $D=3$
\cite{verstraete:comp-power-of-peps,schuch2012resonating}. While the bond
dimension required for $p$ half Trotter steps -- 
corresponding to singlet coverings which contain long-range singlets with
range $p+1$ -- grows as $D_p=3\times 2^{p-1}$,
 a number $p$ of steps (and thus a
singlet span) large enough to yield competitive energies for the
anisotropic limit can be reached with computationally accessible bond
dimension. A key advantage of this explicit PEPS construction, which is
obtained from the RVB PEPS by applying cooling steps, is that it gives us
direct access to the relevant entanglement properties for determining the
topological nature of the system, and thus allow for a direct and
unambiguous identification of the quantum phase of the wavefunction.

The outline of this paper is as follows. In Sec.~\ref{sec:simplexrvb} we
motivate and formally define the simplex RVB ansatz, and give its  PEPS
construction.  In Sec.~\ref{sec:results}, we present our numerical
results: First, we discuss the optimal variational energies  and
corresponding parameters for our ansatz; second, we use PEPS techniques
for analyzing the quantum phase of the system as well as the properties 
of its topological (anyonic) excitations (specifically, anyon masses and
order parameters for anyon condensation and deconfinement); and third, we
discuss the physical structure of the optimal wavefunction (this is 
possible due to the clear physical picture behind our variational
parameters), as well as possible extensions to potentially further improve the ansatz.
Finally, we summarize our results and give an outlook in Sec.~\ref{sec:summary}.

\section{Simplex RVB ansatz and PEPS}\label{sec:simplexrvb}

The key idea behind the simplex RVB ansatz is motivated by the
construction used in Ref.~\onlinecite{poilblanc2013simplex} to study the
isotropic KHAFM.  The motivation behind it is to transform a quantum state
into the ground state of a given Hamiltonian by algorithmic cooling, that
is, by applying local quantum gates (or potentially more broadly local
modifications to the wavefunction) which systematically lower the energy
where applied.
Unlike the construction of the GS in terms of applications of a
trotterized imaginary time evolution operator, where all evolution
operators are chosen identical and close to the identity, in the simplex
ansatz the step size of each evolution operator is optimized
variationally such as to minimize
the energy. In addition, we start from a well-chosen initial state
which already by itself captures essential features of the low-energy
physics of the system at hand.

For what follows, it will be convenient to rewrite the Hamiltonian~\eqref{eq:heis1} as
\begin{equation}\label{eq:hamiltonian}
	H = \tfrac{3}{2} \left( \sum_{ (ijk) \in \triangleright }\!\!\!\!{P_{(ijk)}} +   \gamma 
	\!\!\!\sum_{ (ijk) \in \triangleleft }\!\!\!\!{P_{(ijk)}} \right) - 
	\tfrac{3(1+\gamma)}{8} \sum_{(ijk)}{\mathbb{I}}\ ,
\end{equation}
where $P_{(ijk)}$ is a projector onto the spin-3/2 subspace of 
$\tfrac{1}{2}\otimes\tfrac12\otimes\tfrac12$.
In order to obtain an approximation of the ground state of $H$, we can perform imaginary time evolution $\ket{\psi}=e^{-\beta H}\ket{\phi_\mathrm{init}}$ for sufficiently large $\beta$ and a suitable initial state $\ket{\phi_\mathrm{init}}$. If we trotterize $e^{-\beta H}$, this yields 
\begin{equation}
\label{eq:trotter}
\ket\psi = 
    \left[Q^{\triangleleft}(\alpha_\triangleleft)
        Q^{\triangleright}(\alpha_\triangleright)
        \right]^{K}\ket{\phi_\mathrm{init}}
\end{equation}
with 
\begin{equation}
	Q^{\triangleright}(\alpha_\triangleright):=
	        \prod_{(ijk)\in\triangleright}{ \left( \mathbb{I} 
	        -\alpha_\triangleright P_{(ijk)} \right)}\ ,
\end{equation}
and accordingly for $Q^{\triangleleft}(\alpha)$.

When applied to a suitable initial state, such as the nearest-neighbor RVB
state (i.e., a superposition of all nearest-neighbor singlet coverings
of the lattice), Eq.~\eqref{eq:trotter} has a natural interpretation:
First, it is known~\cite{elser1993kagome,zeng1995quantum} that each NN
singlet covering on the kagome lattice contains exactly 25\% of ``defect
triangles'', that is, triangles which don't contain a singlet
(Fig.~\ref{fig:kagome_1}). 
Those defect triangles have overlap with the spin-$3/2$ subspace and thus
incur a higher energy than triangles holding a singlet (whose energy is
locally optimal). The effect of $Q^{\triangleright}$ ($Q^{\triangleleft}$) is to
decrease the weight of defect triangles on right-pointing (left-pointing)
triangles.  This is achieved by creating longer-range singlets: Since
\begin{equation}
\label{eq:perm-is-rotate}
	P_{(ijk)} = \frac{1}{3} \{\mathbb{I} + R_{(ijk)} + R_{(ijk)}^2 \}
\end{equation} 
(where $R_{(ijk)}$ rotates the qubits), 
acting with $P_{(ijk)}$ on a defect triangle produces longer-range
singlets, i.e.
\begin{equation}\label{eq:P3_half_action}
    \cbox{21.0}{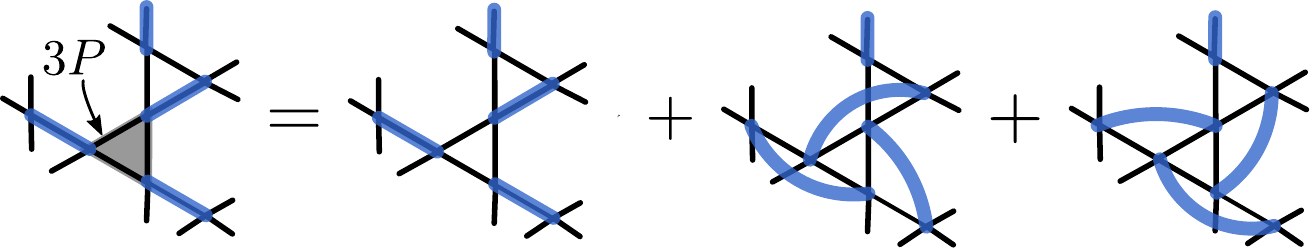} \nonumber\quad .
\end{equation}
(Note, however, that the linear dependence between pairwise permutations and rotations for $3$ qubits -- which allows for the form \eqref{eq:perm-is-rotate} -- implies that this way of looking at the longer-range singlet patterns is not unique; a unique pattern within the singlet subspace can be singled out by avoiding crossings.)

Unfortunately, representing imaginary time evolution accurately using
a Trotter expansion such as 
Eq.~\eqref{eq:trotter} is costly, as it requires a large number of Trotter
steps which grows with the system size; in particular, in the context of
tensor networks this incurs an exponentially growing bond dimension.  One
option here is to compress $e^{-\beta H}$ to a more compact tensor
network~\cite{hastings:locally,molnar:thermal-peps}; however, in this
process, translational invariance is either lost or obfuscated, and the
physical interpretation of the parameters 
missing. We therefore resort to
a different approach: We restrict to a small number of ``Trotter'' layers
in Eq.~\eqref{eq:trotter}, but we allow for independent parameters
$\alpha_i$
for each step $i$ and optimize all those parameters such as to minimize the
variational energy. This leads to the following ansatz, which we term
\emph{simplex RVB}:
\begin{equation}\label{eq:simplex-ansatz}
	\left| \text{RVB}(\bm{\alpha})\right\rangle = 
    Q^{\triangleright}(\alpha_1) 
	Q^{\triangleleft}(\alpha_2)\cdots Q^{*}(\alpha_p) \left| \text{NN RVB}\right\rangle,
\end{equation}	
where $*\in \{\triangleright, \triangleleft\}$ is determined by the parity
of $p$, and the $\alpha_i$, $i=1,\dots,p$, are the variational parameters.
Note that we choose to apply  $Q^\triangleright$ leftmost: This way, the ansatz 
yields the correct behavior in leading order perturbation theory around $\gamma=0$ 
(we discuss this in detail in Sec.~\ref{sec:results_energies}); in agreement with this, we 
observe that this ordering gives better energies in particular in the limit of strong anisotropy.

\begin{figure}[!t]
	\centering
	\includegraphics[width=0.95\columnwidth]{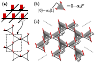}
	\caption{Construction of the tensor network for the simplex RVB. 
	\textbf{(a)} Each local tensor contains three  kagome spins. In particular, the NN RVB has a PEPS description of this form with bond dimension $D=3$.	
	\textbf{(b)} The operator $Q=\mathbb I-\alpha_i P$ acts on three physical spins. It can be considered as an operation controlled by a control qubit in state $\ket0-\alpha_i\ket1$.
	\textbf{(c)}~The tensor network for the simplex RVB for $p=3$, obtained by 
	three applications of $Q$.}
	\label{fig:kagome}
\end{figure}

We now give the PEPS description of the simplex RVB ansatz. We start by
reviewing the construction of the NN RVB
state~\cite{verstraete:comp-power-of-peps,schuch2012resonating} which is
comprised of triangular and on-site tensors. The triangular tensor is
defined to be the sum of one configuration with a defect (containing no singlet)
and three configurations without defect (containing one singlet each):
\begin{equation}\label{eq:eps}
	\cbox{2.5}{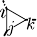}\ =  \delta_{i2}\delta_{j2}\delta_{k2} +
\varepsilon_{ijk}\ ,
\end{equation}
where $\delta$ and $\varepsilon$ denote 3-dimensional Kronecker delta and
fully antisymmetric tensors, respectively. The on-site tensor ensures that every site is paired with exactly one of its neighbors,
\begin{equation}\label{eq:P}
	\cbox{2.5}{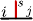}\ = \left( \delta_{i0}\delta_{s0} + \delta_{i1}\delta_{s1}\right)\delta_{j2} + 
	\left( \delta_{j0}\delta_{s0} +
\delta_{j1}\delta_{s1}\right)\delta_{i2}\ .
\end{equation}
The resulting tensor network, obtained by blocking the triangular and
on-site tensors, has a three-site unit cell and is given in
Fig.~\ref{fig:kagome}a.  We implement each local action
$(\mathbb{I}-\alpha P)$, which is not unitary, as a ``controlled'' gate on
three qubits, controlled by a control qubit (this will be useful for extensions of the ansatz discussed
later). The gate acts trivially when the control qubit is $\vert0\rangle$,
while a projector onto the energetically favorable 
spin-1/2 subspace of the three qubits is applied if the control qubit is
$\vert1\rangle$ (Fig.~\ref{fig:kagome}b). For the time being, we choose
the control qubits in a product state $\vert0\rangle + \alpha\vert1\rangle$,
leading to a gate $Q=(\mathbb{I} - \alpha P)$, as described previously.
For illustration, the tensor network obtained through three applications of $Q$ to the NN RVB,
starting with the right-pointing triangles,  
is shown in Fig.~\ref{fig:kagome}c. 

What is the bond dimension of the simplex RVB PEPS with $p$ layers? We
work with the square unit cell shown in
Fig.~\ref{fig:kagome}a, which contains three kagome spins. With this unit
cell and the triangular tensor (\ref{eq:eps}), the NN 
RVB itself has $D=3$.  $Q^\triangleright(\alpha_i)$ on right-pointing triangles lie within the unit cell
and therefore carry no increase in the bond dimension.
Operators
$Q^\triangleleft(\alpha_i)$ on left-pointing
triangles can be implemented with bond dimension $4$: E.g., they can be
constructed  by teleporting the left and bottom neighboring spin to the
central site 
(cf.~Fig.~\ref{fig:kagome}a),
applying $Q^\triangleleft(\alpha_i)$, and teleporting them back. $D_p$ is therefore
multiplied by $4$ for every even $p$, i.e., $D_p=3,12,12,48,\dots$ for
$p=1,2,3,4,\dots$\;.  However, for $p$ even we can do better: 
There, the rightmost $Q^\triangleleft$ is applied directly to the NN RVB, 
in which 
case the state of the teleported spins is already known to the central
tensor if the NN RVB index is $0$ or $1$, allowing to compress the bond
dimension for $p=2$ to $D_2=6$;  thus, we obtain 
$D_p=3\times 2^{p-1}=3,6,12,24,\dots$\;\footnote{The same compression can also be obtained with the opposite blocking, noting that in expectation values, the final $Q^\triangleright$ appears as 
${Q^\triangleright(\alpha_i)}^\dagger Q^\triangleright(\alpha_i)\propto Q^\triangleright(2\alpha_i-\alpha_i^2)$, which can be implemented with bond dimension $4$, i.e.\ $2$ per ket/bra layer.}.

\section{Results}\label{sec:results}

Let us now discuss our results obtained by using simplex RVB states as a
variational ansatz for the breathing kagome Heisenberg Hamiltonian \eqref{eq:heis1}. The
PEPS formalism enables the computation of expectation values of local
observables and correlation functions directly in the thermodynamic limit,
in contrast to other methods.  We use standard numerical methods for
infinite PEPS (iPEPS)~\cite{jordan:iPEPS,haegeman2017diagonalizing}, which
approximate the boundary by an infinite matrix product state (iMPS) of bond
dimension $\chi$ (which determines accuracy and computational cost).  This
allows us to compute the variational energy of an iPEPS with high accuracy
and thus to determine the variationally optimal state.  In addition, the
PEPS approach allows us to utilize the entanglement symmetries of the PEPS
and the way in which the iMPS boundary orders relative to those symmetries
to study the quantum phase and the topological properties of the optimized
wavefunction~\cite{duivenvoorden2017entanglement,iqbal2018study}.  In our
calculations, we choose not to truncate the PEPS tensor before contraction
but rather keep the exact simplex ansatz, which avoids truncation errors
and gives us direct access to the entanglement symmetries relevant to
study the nature of the order in the system; on the other hand, this limits our ansatz to
at most $p=3$ computationally attainable applications of $Q$'s.

\subsection{Energies\label{sec:results_energies}}

Let us start by giving the results on the optimal variational energy
obtained within the simplex RVB ansatz family.  For all calculations, we
have determined the optimal parameters $\{\alpha_i\}$ through a gradient
search using the corner transfer matrix method
with a boundary bond dimension $\chi = 36$,
and subsequently extracted the energies of the optimized
wavefunctions using boundary iMPS  (i.e., the fixed point of the transfer
operator) with $\chi=64$  (this only requires truncation along one
direction,
resulting in a better convergence of the energy in $\chi$).  A
table with the detailed energies, including a convergence analysis and
error bounds, as well as a discussion of a potential
extrapolation in $p$, are given in Appendix~\ref{app:energy-densities}.

\begin{figure}[t]
	\centering
	\includegraphics[width=27em]{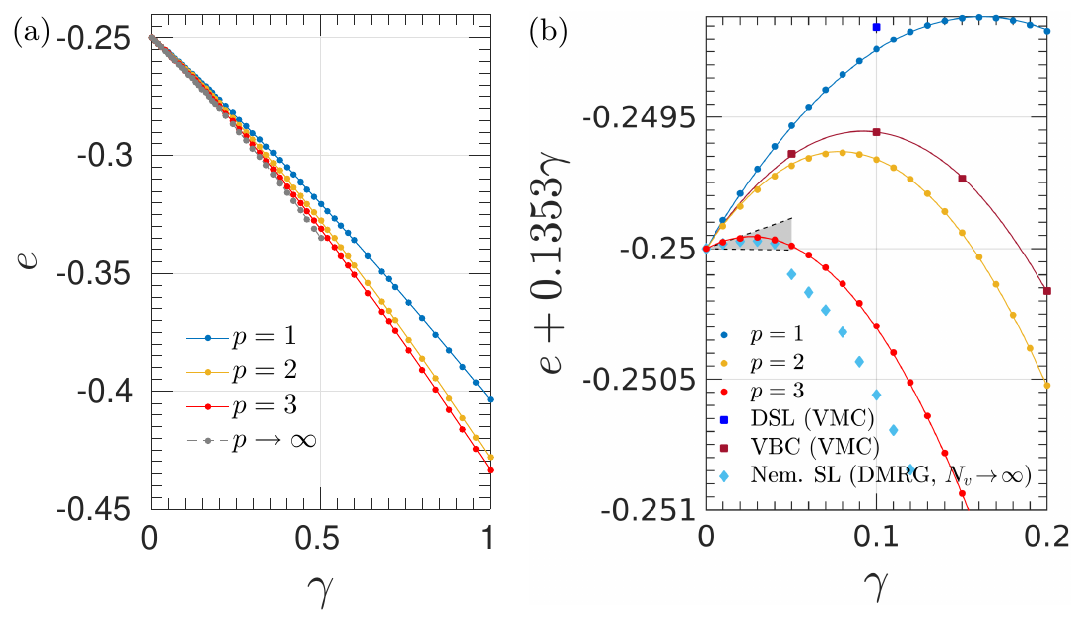}
	\caption{ \textbf{(a)} Energy densities for the simplex RVB ansatz with $p=1,2,3$, and extrapolated values for $p\rightarrow \infty$ (cf.\ Appendix~\ref{app:c1-coefficient}). \textbf{(b)} Comparison of energy densities for the simplex RVB ansatz with the data from variational Monte Carlo (VMC) \cite{iqbal2018persistence} and DMRG \cite{repellin2017stability, cecile:private}. Solid lines give quadratic fits to the data for $\gamma \in [0,0.3]$. The gray region is bounded by lines $-0.25-0.1354\gamma$ and $-0.25-0.133\gamma$, which are the slopes extracted from DMRG calculations for the full model for $N_v=12$ and the extrapolation $N_v\to\infty$, respectively.}
	\label{fig:main_energy}
\end{figure}

In Fig.~\ref{fig:main_energy}a, we plot the energy density $e$ (i.e., the energy per site) of the
optimized simplex RVB wavefunction for the breathing kagome
Hamiltonian~\eqref{eq:heis1} as a function of the anisotropy $\gamma$, for
$p=1,2,3$.  
For better comparison in the strongly anisotropic limit, we plot 
in Fig.~\ref{fig:main_energy}b
$e(\gamma)+0.1353\,\gamma$ for $0\le\gamma\le0.2$,
where the subtracted linear offset corresponds to the behavior in first order perturbation theory, as
obtained by extrapolating DMRG calculations~\cite{repellin2017stability,cecile:private}
for the effective first-order model~\cite{mila:breathing}
on finite cylinders. Beyond the $p=1,2,3$ simplex RVB results,
we also show 
the data obtained for the VBC (and the energetically less favorable
$\mathrm{U}(1)$ DSL) ansatz in Ref.~\onlinecite{iqbal2018persistence} using VMC. 
We find that already for $p=2$,  our ansatz gives energies
slightly below the VBC ansatz, and for $p=3$, it clearly outperforms it.
This is particularly remarkable since the $p=2$ ($p=3$) simplex ansatz has
only two (three) parameters, corresponding to effectively one (one and a
half) imaginary time evolution steps, while the VMC ansatz of
Ref.~\onlinecite{iqbal2018persistence} has $11$
parameters, including two Lanczos steps
(cf.\ the discussion in the introduction). 
In addition, Fig.~\ref{fig:main_energy}b also shows energies obtained by extrapolating DMRG data for the full model \eqref{eq:heis1} for cylinders with $N_v=8,10,12$ to $N_v\to\infty$, which we find to be remarkably close to our $p=3$ data in the strong anisotropy regime around $\gamma\le0.04$, given that our ansatz only depends on three parameters rather than about $10^8$. Since the extrapolation of the DMRG data is subtle (cf.\ Appendix~\ref{app:c1-coefficient}), the gray cone indicates the linear order extracted from the $N_v=12$ and $N_v\to\infty$ DMRG data, which we expect to provide reliable lower and upper bounds to the true slope for the full model.

For better comparison in the strong anisotropy limit, 
we expand the energy density for small $\gamma$ as 
\begin{equation}
    e(\gamma)=-0.25 + c_1\gamma+ c_2\gamma^2 +\dots \, ,
\end{equation}
where the $c_i$ depend on $p$. 
The values $c_1$ for the slope
at $\gamma=0$  for the different methods are given in Table~\ref{table:c1}
(see Appendix~\ref{app:c1-coefficient} for details on the extraction).
They confirm
that for $p=2$, our ansatz performs slightly better than the VBC
ansatz~\cite{iqbal2018persistence}, while for $p=3$, it clearly
outperforms it.  The DMRG data for the nematic spin liquid is the same as the gray cone in Fig.~\ref{fig:main_energy}b, that is, obtained from the DMRG data of
Ref.~\onlinecite{repellin2017stability}~\cite{cecile:private} for $N_v=12$ cylinders and the $N_v\to\infty$ extrapolation, which should give lower and upper bounds to the true value.
Finally, we give values for $c_1$ obtained by extrapolating to $p\to\infty$ in the inverse bond dimension $1/D_p\sim 1/2^p$, which we expect to be a reasonable fit in a gapped phase, and which yields a value competitive with the DMRG results.

\begin {table}[t]
\begin{center}
	\bgroup
	\def\arraystretch{1.4}%
	\begin{tabular}{r @{\ }||@{\ } l  l} 
		Ansatz \hspace*{3em} &  \hspace*{1.5em}$c_1$ \\ 
		\hline\hline
		U(1) DSL (VMC) \cite{iqbal2018persistence} & $-0.119(1)$    \\ \hline
		VBC (VMC) \cite{iqbal2018persistence} 	 & $-0.125545(20)$  \\ \hline
		Nematic SL (DMRG)
		& 
		$-0.1354$ & (fit, $N_v=12$) \\ 
		\cite{repellin2017stability,cecile:private} 
		   & $-0.133$ &(fit, $N_v\!\to\!\infty$)  \\ \hline
		simplex RVB, $p=1$ 	 & $-0.1242$ &(fit)
		\\
		 & $-0.1241978(4)$ & ($e^\star_\triangleleft$)\\
		 & $-0.1243(3)$ & (\cite{iqbal2018persistence}, extrap.\ $N_v$) \\ \hline
		\dots, $p=2$  				 & $-0.1261$ & (fit) \\
		& $-0.126217(7)$ & ($e^{\star}_{\triangleleft}$)
		\\\hline
		\dots, $p=3$  			 & $-0.1319$ & (fit) \\
		 & $-0.13225(5)$
		& ($e^{\star}_{\triangleleft}$)   \\\hline
		\dots, $p=\infty$  					 &  $-0.1345$& (extrap.\ fit) \\ 
		 					 &  $-0.1349$ &(extrap.\ $e^\star_\triangleleft$)  
	\end{tabular}
	\egroup
\end{center}
\caption{
\label{table:c1}
Coefficient of the linear term in the energy density $e\approx -0.25 +c_1\gamma$, 
obtained with different methods (see text). The simplex RVB values labelled $e^\star_\triangleleft$ have been obtained using first-order perturbation theory, cf.~text.}
\end {table}

An alternative way to extract $c_1$ is by using a perturbative expansion.  Perturbation theory predicts that for small $\gamma$, the energy per site is given by
\begin{equation}
e(\gamma)\approx -0.25+c_1\gamma=\frac1N \min_{\ket\psi\in\mathcal G} \bra\psi H(\gamma)\ket{\psi}\ ,
\end{equation}
where $\mathcal G$ denotes the ground state manifold at $\gamma=0$, that is, the subspace of all states with spin $\frac12$ on the right-pointing triangles. Within our variational family, this corresponds to fixing $\alpha_1=1$ (which explains why we want to have $Q^\triangleright$ leftmost in Eq.~(\ref{eq:simplex-ansatz}) if we want to correctly reproduce the perturbative limit), and letting $\mathcal G$ be the set of simplex RVBs for a given $p$ with fixed $\alpha_1=1$. We thus find that within our ansatz family, we can determine $c_1$ perturbatively as
\begin{align*}
c_1 &= \frac1\gamma
\left[\frac1N
\min_{\ket\psi\in\mathcal G} \bra\psi 
\textstyle\sum_{\triangleright }{\textbf{S}_i . \textbf{S}_j}
+  \gamma \textstyle\sum_{\triangleleft }{\textbf{S}_i . \textbf{S}_j}
\ket\psi  + 0.25
\right]
\\
&=\frac1N\min_{\ket\psi\in\mathcal G} \bra\psi 
\textstyle\sum_{\triangleleft }{\textbf{S}_i . \textbf{S}_j}
\ket\psi \ ,
\end{align*}
that is, by minimizing the energy density $e^\star_\triangleleft$ on the left-pointing triangles within the simplex RVB family with $\alpha_1=1$. This optimization incurs one less variational parameter and does not require fitting, and can thus be carried out to significantly higher precision; we report the corresponding values in Table~\ref{table:c1} alongside the values obtained from fitting $e(\gamma)$, which are in excellent agreement.

\begin{figure}[!t]
	\centering
	\includegraphics[width=26em]{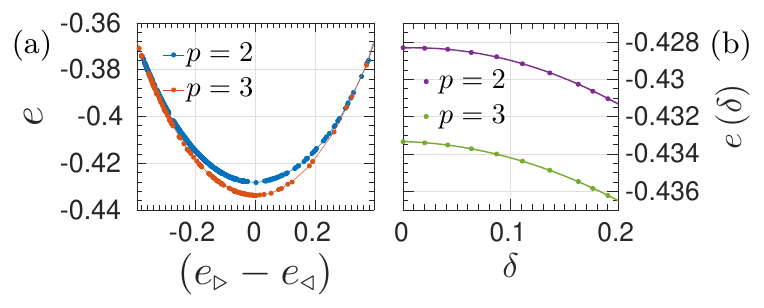}
	\caption{ 
    Restored inversion symmetry at the Heisenberg point.  \textbf{(a)} Convex hull of energy densities for $\gamma=1$ vs.\ energy difference between left- and right-pointing triangles for the $p=2,3$ simplex ansatz states.  The inversion symmetry, which is not explicitly contained in the ansatz, is essentially perfectly restored in the energy.
	 \textbf{(b)} Optimal energies for $p=2,3$ in the symmetric gauge, where $\delta$ measures the distance to the  Heisenberg point, with  quadratic fits. The slope at $\delta=0$ is essentially zero, 
	reconfirming that inversion symmetry of the energy is restored at the Heisenberg point. }
	\label{fig:leftright}
\end{figure}

As an additional check for the quality of the optimal variational state,
we have considered the energetics of left- and right-pointing triangles at
and around the Heisenberg point.  We find that even though the $p=3$
ansatz treats the inequivalent triangles differently (in particular, $Q$
acts twice on the the right- and only once on the left-pointing
triangles), the energy splitting between the triangles vanishes for the
optimal energy wavefunction (Fig.~\ref{fig:leftright}a); we observe the
same effect also for the optimal wavefunction with $p=2$. Alternatively,
we can consider the optimal energy density in a symmetric gauge, 
$H(\delta) =  (1-\delta)\sum_{ \triangleright }{\textbf{S}_i . \textbf{S}_j}
+ (1+\delta) \sum_{ \triangleleft }{\textbf{S}_i . \textbf{S}_j}$, in the vicinity of the Heisenberg point. We obtain a fit 
$e^{p=2}_{\mathrm{GS}}(\delta) \approx 
- 0.4283 +0.001\delta -0.083\delta^2$ and 
$e^{p=3}_{\mathrm{GS}}(\delta) \approx 
- 0.4333 -0.001\delta -0.071\delta^2 $ 
for the simplex RVB ansatz with
$p=2$ and $p=3$, respectively (Fig.~\ref{fig:leftright}b), which
essentially show a zero  slope at $\delta=0$ and is thus
symmetric around $\delta=0$ to very good accuracy, as required by symmetry
considerations~\footnote{In fact,
the two tests probe the same property: Equal energies for both triangles
are equivalent to zero slope at $\delta=0$, since otherwise, the
wavefunction for some non-zero $\delta$ would constitute a better ansatz
even at $\delta=0$.}.

\subsection{Order, correlations, and quantum phase}\label{ssec:qph}

The PEPS description of the NN RVB has a \z{2} symmetry on the
entanglement degrees of freedom.  Such a symmetry has been shown to be
essential to explain the topological features of PEPS models, as well
as to understand and analyze the breakdown of topological order in such
systems~\cite{chen:topo-symmetry-conditions,schuch:peps-sym,schuch2013topological,haegeman2015shadows,duivenvoorden2017entanglement,iqbal2018study}.
This is accomplished by considering the boundary iMPS obtained when
contracting the 2D PEPS (i.e., the fixed point of the transfer operator),
and analyzing how it orders relative to those symmetries: The specific
type of order is directly related to the quantum phase displayed by the
bulk wavefunction.  
From those symmetries, we can construct half-infinite string operators which on the one hand 
create anyonic bulk excitations in a given anyon sector $a$ (with
$a=s,v,sv$ for spinon, vison, and the composite fermion, respectively),
but at the same time form (string) order parameters which detect the
ordering of the boundary state.  By computing expectations values of these
string operators either in one layer (denoted $\langle a\rangle$) or in
both layers (denoted $\langle a a^\dagger\rangle$), we can construct order
parameters which probe the condensation and confinement of anyons, and
thus the proximity to a topological phase transition; specifically, a
non-zero value of the ``deconfinement fractions'' $\langle
aa^\dagger\rangle$, as well as ``condensate fractions'' $\langle
a\rangle\equiv 0$, are indicative of the topological phase.  At the same
time, for vanishing order parameters we can study the rate at which the
corresponding expectation value for finite strings with two dual endpoints
decays to zero as their separation increases, giving rise to corresponding
length scales for condensation (mass gap) and confinement.

\begin{figure}[t]
	\centering
	\includegraphics[width=27em]{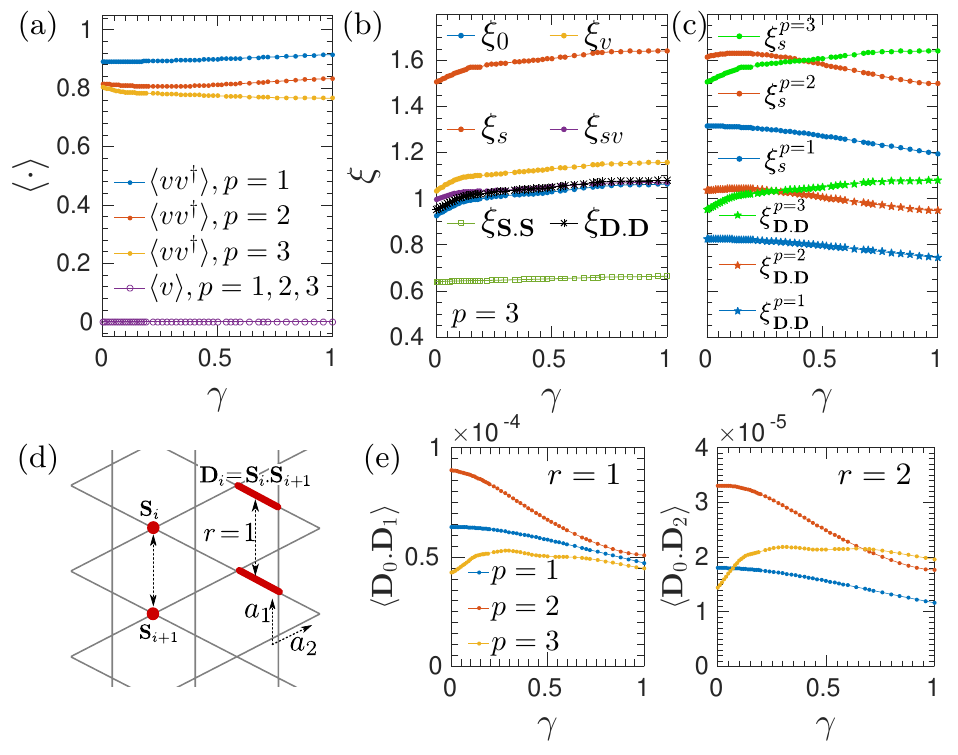}
	\caption{\textbf{(a)} Deconfinement fraction $\langle vv^{\dagger} \rangle$ and condensate 
	fraction $\langle v \rangle$ of visons as a function of $\gamma$. 
	\textbf{(b)} Different correlation lengths for $p=3$, computed with $\chi=192$: 
    For the trivial ($0$), spinon ($s$), vison ($v$), and fermionic ($sv$) sector, as well as the spin-spin ($\mathbf S.\mathbf S$) and dimer-dimer ($\mathbf D.\mathbf D$) correlations shown in (d).
	\textbf{(c)}~Comparison of $\xi_s$  and $\xi_{\mathbf{D}.\mathbf{D}}$
	for $p=1,2,3$. 
	\textbf{(d)} Spin-spin and dimer-dimer correlation functions considered in (b,c,e). \textbf{(e)} $\langle \mathbf{D}_i.\mathbf{D}_{i+r} \rangle$ for $r=1,2$ and $p=1,2,3$.}
\label{fig:cf_gamma}
\end{figure}

We have computed anyonic order parameters for the vison, as well as
correlation lengths for all anyons, for the optimized simplex RVB for
$p=1,2,3$ as a function of the anisotropy $\gamma$.  Since we do not
truncate local tensors during optimization, the entanglement symmetries of
our tensors remain easily accessible, facilitating the analysis.   The
corresponding data is shown in Fig.~\ref{fig:cf_gamma}.  For the anyonic
order parameters (Fig.~\ref{fig:cf_gamma}a), we find that $\langle
v\rangle=0$ and $\langle vv^\dagger\rangle>0$, which implies that the
system is in a $\mathbb Z_2$ topologically ordered phase for the given $p=1,2,3$. The 
$\langle vv^\dagger\rangle$ for the different $p$ all show only a small $\gamma$-dependence, with no indication of a phase transition at some intermediate $\gamma$ building up at larger $p$.  On the other hand, at least for $\gamma$ close to $1$, $\langle vv^\dagger\rangle$ clearly decreases with $p$, leaving open the possibility of a critical phase around the Heisenberg point.

Next, let us analyze the correlation lengths,  shown in
Fig.~\ref{fig:cf_gamma}b for $p=3$.  We find that the dominant correlation
length is given by spinon correlations, as known for the NN RVB
state~\cite{haegeman2015shadows}.  In addition to the different anyon
correlations, the figure also shows data for spin-spin 
($\langle \mathbf{S}_i.\mathbf{S}_{i+r} \rangle$) 
and dimer-dimer 
($\langle \mathbf{D}_i.\mathbf{D}_{i+r} \rangle$)
correlations, 
computed for the spin and dimer pairs indicated in
Fig.~\ref{fig:cf_gamma}d.
Again, all lengths change smoothly with $\gamma$, and while we observe a minor increase 
of correlations with $\gamma$, there is no sign of a phase transition. Note that the similar behavior of spinon and leading trivial (including dimer-dimer) correlations and their relative scale is consistent with previous observations~\cite{iqbal:rvb-perturb} which could be explained as arising from correlations between pairs of spinons~\footnote{It is worth noting that the values we obtain
for $\xi_s$ and $\xi_v$ for the NN RVB wavefunction are in remarkable
agreement with their earlier estimates in
Ref.~\onlinecite{poilblanc2013simplex} which had been extracted from the
finite-size scaling of the energy splitting for the corresponding ground
states.}.

In Fig.~\ref{fig:cf_gamma}c, we compare the spinon and dimer
correlation lengths $\xi_\bullet^p$ for the different $p=1,2,3$. 
We find a surprising behavior: While the curves for $p=1$ and $p=2$ show qualitatively similar behavior (with increased correlation length for $p=2$), and display a decrease of correlations with growing $\gamma$, the $p=3$ curve exhibits the opposite behavior.  Even more noteworthily, while at the Heisenberg point, correlations keep increasing with $p$ (consistent with a gapless phase), in the small $\gamma$ regime the correlations decrease again, speaking against a long-range ordered or critical system.  
The behavior for $p=1,2$ can be qualitatively explained from the 
way the $Q(\alpha)$ act (cf.\ the next section where the optimal
$\alpha$ are discussed): $Q^\triangleright(\alpha_1)$ acts on
the strong right triangles. As it decreases the energy of the latter,
$\alpha_1$ will increase with growing anisotropy. At the same time, the
$Q$'s create longer-range singlets, which should give rise to longer-range
correlations: Correlation functions are obtained from overlaps of singlet
configurations, weighted by the number $\ell$ of singlets involved as
$2^{-\ell/2}$, and longer-range singlets allow to connect two points at a
given distance with
smaller $\ell$ and
thus larger weight~\footnote{Note that this is only a qualitative argument, as it
does not take into accounts cancellations due to different phases, or the
scaling of the number of loop patterns with the distance.}.
The additional increase in correlation length for $p=2$ can be explained from the presence of the $Q^\triangleleft(\alpha_2)$ layer which gives rise to additional longer range singlets before the application of $Q^\triangleright(\alpha_1)$.
But why does the behavior change for $p=3$?  As we discuss in more detail in the next section, the role of the topmost $Q$'s is to adjust the energy, as they shift the weight between spin $\frac12$ and $\tfrac32$ right before applying the Hamiltonian; the optimal value of the corresponding $\alpha_i$ -- and thus the amount of correlations they create -- is thus governed by immediate energetic considerations (i.e., the overlap with the spin $\tfrac12$ space).  The lower-lying $Q$ layers, on the other hand, are not directly relevant for the energetics -- rather, their job is to set up the underlying wavefunction by creating longer-range singlets in a way where the topmost layers can produce the best possible energies for left- and right-pointing triangles simultaneously. Thus, it is only with the lower layers $i\ge3$ that the $Q$'s primarily serve the purpose of creating the right type of long-range singlets and correlations, rather than just tuning the value of the energy.   It therefore seems plausible that the $p=3$ behavior of the correlations is closer to the true behavior at large $p$, and 
we expect this tendency to continue as we further increase $p$.

Finally, let us discuss the possibility of a nematically ordered phase, as
proposed in Ref.~\cite{repellin2017stability} for the strongly anisotropic
limit; here, the nematic order was found to break rotational symmetry
around the center of the triangles, leading to different Heisenberg
energies along inequivalent links.  By construction, our ansatz keeps all
symmetries of the Hamiltonian and thus cannot explicitly break this
symmetry. On the other hand, if nematic order were favored we would expect
the system to form a long-range ordered state, that is, an equal weight
superposition of all three nematically ordered states. This long-range
order should be reflected in a diverging correlation length, and thus, the
absence  of any such divergence 
in the dimer-dimer correlations (which in fact rather decrease) speaks against the
presence of a nematically ordered phase. To further strengthen this point,
we have also considered the dimer-dimer correlations $\langle \bm D_0.\bm
D_r\rangle$ at a given short distance $r$, shown in
Fig.~\ref{fig:cf_gamma}e -- if nematic order were favorable, we
would expect to see such correlations build up at short distance already at low $p$.
However, Fig.~\ref{fig:cf_gamma}e shows no significant increase of $\langle \bm
D_0.\bm D_r\rangle$ at small $\gamma$, and it in fact decreases for $p=3$. 
More importantly, the observed values of $\langle \bm D_0.\bm
D_r\rangle$ are of the order of $10^{-4}$ even for $r=1$, while for the
nematic order  parameter found in Ref.~\cite{repellin2017stability}, we would expect
them to be on the order of $0.03$. Overall, we find that our results show no
indications of a nematic phase in the strong anisotropy limit $\gamma\ll
1$.

\begin{figure}[t]
	\centering
	\includegraphics[width=24em]{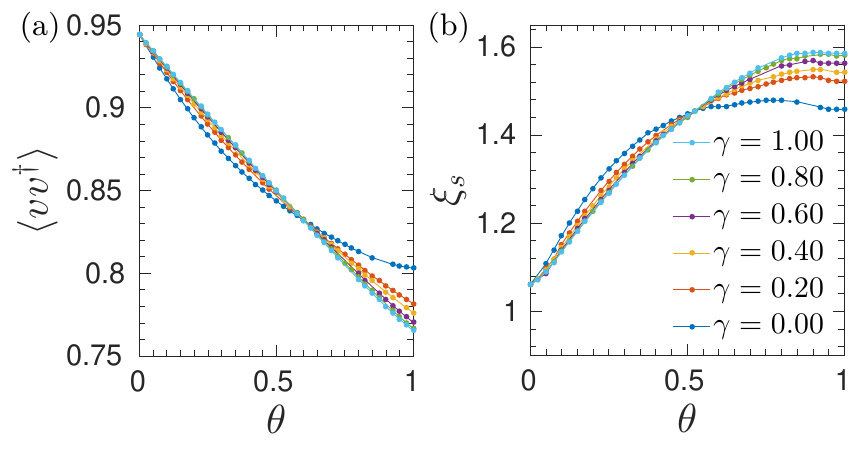}
	\caption{ Curves in (a) and (b) share the same color for each $\gamma$ along the path $\bm{\alpha}(\theta) = \theta\times \bm{\alpha^{\star}}_\gamma$ for the $p=3$ simplex ansatz. (a) The deconfinement fraction of visons. (b) The length scale of spinon excitations with $\chi=144$.}
	\label{fig:interpolation}
\end{figure}

As a final test for the $\mathbb Z_2$ topological spin liquid nature of the simplex RVB state for $p=3$ for all values of $\gamma$, we study the deconfinement order parameter for visons and the spinon correlation length as we interpolate from the NN RVB state (which is known to be a gapped $\mathbb Z_2$ topological spin liquid) to the optimal simplex RVB wavefunction, by interpolating
$ \bm{\alpha}(\theta)=\theta \times \bm{\alpha^{\star}}_\gamma$ 
from $\theta=0$ (NN RVB)  to $\theta=1$ (optimized simplex RVB), 
where $\bm{\alpha^{\star}}_\gamma$ denotes the optimal parameter values for a given $\gamma$. 
The result is shown in Fig.~\ref{fig:interpolation}.
Again, we find that both quantities change smoothly, re-confirming the topological $\mathbb Z_2$ spin liquid nature of the optimal wavefunction for the whole range of $\gamma$.

\subsection{Structure of optimal wavefunction \mbox{and possible generalizations}}

The fact that our simplex RVB ansatz encompasses only very few parameters with
a clear interpretation allows us to directly study how the structure of
the optimal wavefunction changes as we vary $\gamma$ and increase $p$.
We recall that our ansatz [Eq.~\eqref{eq:simplex-ansatz}] was of the form
\begin{equation}
	\left| \text{RVB}(\bm{\alpha})\right\rangle = 
    Q^{\triangleright}(\alpha_1) 
	Q^{\triangleleft}(\alpha_2)\cdots Q^{*}(\alpha_p) \left| \text{NN RVB}\right\rangle,
\end{equation}	
where $Q^{\bullet}(\alpha_i)$, $\bullet\in\{\triangleleft,\triangleright\}$ projects
onto the spin-$1/2$ subspace of the corresponding triangles for
$\alpha_i=1$ and acts trivially for $\alpha_i=0$ -- that is, it lowers the
energy of those triangles as $\alpha_i$ approaches $1$.  At the same time,
it increases the number of longer-range singlets, as it acts by permuting
the singlets; following Eq.~\eqref{eq:perm-is-rotate}, one can argue that the largest amount of singlets is permuted at $\alpha=3$, though this has to be taken with due care due to the large number of linear dependencies of different long-range singlet patterns, as well as cancellations in the singlet range growth from permutations on adjacent sites.  Indeed, the fact that the $Q$'s arise from trotterizing the
imaginary time evolution implies that a certain amount of such long-range
singlets is \emph{required} to obtain a good variational wavefunction.

\begin{figure}[t]
	\centering
	\includegraphics[width=23em]{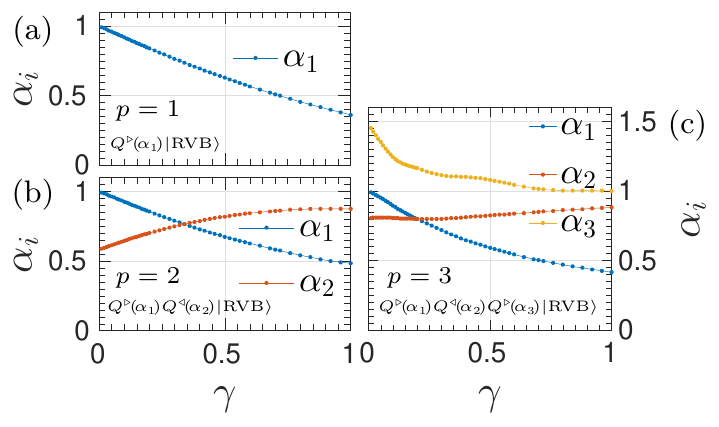}
	\caption{Optimal values of variational parameters for the simplex RVB ansatz for $p=1,2,3$. }
	\label{fig:params_conv}
\end{figure}

Fig.~\ref{fig:params_conv} shows the optimal values of
$\{\alpha_i\}_{i=1}^p$ for $p=1,2,3$, as a function of the breathing
anisotropy parameter $\gamma$. For $p=1$, we find that for 
maximum anisotropy $\gamma=0$, $\alpha_1=1$ -- this is expected, as it
forces all strong triangles to have spin $1/2$ and thus minimum energy,
and the state of the weak triangles is irrelevant for $\gamma=0$.  As we
increase $\gamma$, we observe that the value of $\alpha_1$ decreases,
increasing the probability of the weak triangles to have spin $1/2$.
Remarkably, however, we see that the optimal value of $\alpha_1$ even at
the symmetric point $\gamma=1$ is significantly above $0$: This can be
understood from the fact that $Q$ in the simplex RVB ansatz does act not only 
by
shifting the weights of defects between inequivalent  triangles in the
RVB, but at the same time creates energetically favorable longer-range
singlets. Note, however, that those singlets are created by decreasing, rather than increasing, $\alpha_1$, suggesting that for $p=1$, the amount of longer-range singlets at the Heisenberg point is smaller than for $\gamma\ll1$.

For $p=2$, the behavior of $\alpha_1$ closely resembles that for $p=1$, but with a smaller change (i.e., a larger value $\alpha_1$) towards $\gamma=1$. The correspondingly lower energy gain on the left-pointing triangles is compensated by $Q^\triangleleft(\alpha_2)$, which lowers the energy of the left-pointing triangles prior to the application of $Q^\triangleright(\alpha_1)$, while in addition
allowing for the creation of NNN neighbor singlets, thus being energetically favorable.

For $p=3$, we make a similar observation: The curves for $\alpha_1$ and $\alpha_2$ are again quite close to the 
$p=2$ case. Now, the value of $\alpha_2$ for small $\gamma$ has increased -- giving the weak left-pointing triangles a lower energy, but also creating longer-range singlets.  The energy gain of the weak triangles is now compensated by biasing the system towards strong triangles using $\alpha_3$.  An interesting point to note is that $\alpha_3>1$, unlike $\alpha_1$ and $\alpha_2$: That is, in the first layer applied to the NN RVB, it is now favorable to flip the sign of the spin-$3/2$ space or -- to the extent the picture of Eq.~(\ref{eq:perm-is-rotate}) as creating longer-range singlets is correct -- to create a larger fraction of longer-range singlets at the expense of not lowering the energy of the strong $\triangleright$ triangles as much as possible.  Indeed, the latter interpretation is plausible, given that longer-range singlets are overall energetically favorable, and the immediate energetics 
is taken care of by $Q^\triangleright(\alpha_1)$.  If we were to  follow this reasoning, we would expect further layers to also have $\alpha_i>1$, $i\ge3$; this suggests that the qualitative change in the behavior of order parameters and correlations occurring at $p=3$ will persist for larger values of $p$.

Given the observation that the lowest  layer (i.e., $Q^\bullet(\alpha_p)$) significantly biases the NN RVB towards configurations with no defects on one kind of triangles, 
 it seems plausible that
a modification of the NN RVB layer in a way 
which biases  it 
towards configurations with less defects on the suitable triangles 
should further improve the ansatz. This can be done following the idea of the original simplex RVB paper~\cite{poilblanc2013simplex}: 
We modify the right-pointing triangular tensor
\eqref{eq:eps} of the NN RVB as 
\begin{equation}\label{eq:eps2}
	\cbox{2.5}{eps}\ =  (1-\beta) \delta_{i2}\delta_{j2}\delta_{k2} + \varepsilon_{ijk}\ ,
\end{equation}
where a parameter $\beta>0$ ($\beta<0$) effectively shifts the amplitude of defect configurations towards left-pointing (right-pointing) triangles. Importantly, this modification does not lead to an increase in the bond dimension. 
Subsequently, we can apply the $Q$'s as before,
\begin{equation}\label{eq:simplex-ansatz-2}
    Q^{\triangleright}(\alpha_1) 
	Q^{\triangleleft}(\alpha_2)\cdots Q^{*}(\alpha_p) \left| 
\text{NN RVB} (\beta)\right\rangle\ ,
\end{equation}    
to obtain an enhanced simplex RVB ansatz with one additional parameter.
We have tested this ansatz and found that it does not lead to better variational energies, except in the case $p=1$. We attribute this to two facts:
As discussed, the energetics is predominantly taken care of by the top $Q$ layers (in particular $\alpha_1$ and $\alpha_2$), rather than the low-lying $\beta$; on the other hand,
the lower layers mostly serve the purpose to create longer range singlets, 
whereas $\beta$,
while changing the weight of the spin-$1/2$ space on the corresponding triangles, does not give rise to longer range singlets (and in fact reduces the spinon correlation in the system).

As mentioned earlier, we can consider the 
$Q$ operators as gates which are controlled by a ``control qubit'' $\ket{0}+\alpha_i\ket{1}$. This enables us to interpret the simplex ansatz as a variational optimization over $p$ control qubits chosen from the manifold of product states (Fig.~\ref{fig:kagome}c). In principle, we can further enrich the simplex ansatz \eqref{eq:simplex-ansatz} by allowing the control qubits to be in a general $p$-qubit state, i.e., correlating the $\alpha_i$ of the different layers. This provides a significantly enlarged simplex manifold, even though it does not allow to increase the range of the singlets. However, we have found that for the computationally feasible values of $p$, the optimization over the manifold of general $p$-qubit control states does not lead to an improvement in energy  as compared to the optimization over the manifold of $p$-qubit product states.

\section{Summary and outlook}\label{sec:summary}

In this paper, we have introduced a simple yet powerful ansatz for the kagome Heisenberg antiferromagnet (KHAFM) with breathing anisotropy, termed simplex RVB.  Our ansatz is physically motivated from algorithmic cooling, and effectively consists of $p/2$ imaginary time evolution layers with optimized step sizes applied to the NN RVB, approaching the true ground state as $p\to\infty$. It yields simple few-parameter families of wavefunctions with a clear physical interpretation in terms of longer range singlets, which are energetically favorable. The ansatz has a simple PEPS representation, which makes it amenable to numerical simulations and an in-depth analysis of its order.

We have analyzed the optimal simplex RVB ansatz for $p=1,2,3$ for the breathing KHAFM, with a focus on the strong anisotropy limit, and found that already for $p=2$ it improves over existing VMC results, while for $p=3$ it clearly outperforms them, even though it requires significantly less parameters. We also find that for $p=3$, our energies are rather close to extrapolated DMRG energies. It is thus probable that with just a few more layers, the simplex RVB will be fully competitive with DMRG simulations, which is remarkable given the small number of parameters.

We have investigated the nature of the order in the optimized simplex RVB for the breathing Heisenberg model, using a wide range of probes based on the explicit PEPS description and the underlying entanglement symmetries. We find that for the whole parameter regime, our ansatz yields a gapped $\mathbb Z_2$ topological spin liquid for the accessible values of $p$.  In the strongly anisotropic regime, we find that the correlations saturate as we increase $p$, thus exhibiting no signs of long-range order which one would expect e.g.\ for a nematically ordered phase. On the other hand, at the Heisenberg point, our results show a clear tendency to larger correlation lengths as $p$ increases, which is consistent with a critical DSL phase at the Heisenberg point; both the improvement in energy and the growth of correlations with $p$ points to the relevance of long-range fluctuating singlets for the kagome Heisenberg antiferromagnet.

In order to benchmark a gapped vs.\ a gapless spin liquid in particular at the Heisenberg point, it would be interesting to compare the simplex RVB ansatz with a variant where one starts from a gapless $\mathrm{U}(1)$ spin liquid rather than the gapped NN RVB.  One idea which fits well with the PEPS picture is to change the $\mathbb Z_2$ invariant PEPS tensors by $\mathrm{U}(1)$ invariant ones (which we expect to give a critical wavefunction), e.g.\ by omitting those $\mathbb Z_2$-invariant configurations which break $\mathrm{U}(1)$. We describe and test such an ansatz in Appendix~\ref{app:gaplessrvb}. 
However, while the resulting ansatz indeed yields a gapless spin liquid, we find that the corresponding wavefunction is energetically unfavorable for the Heisenberg model.  The reason for this can be found in the general approach of the construction:
Since different $\mathbb Z_2$ configurations map to different singlet patterns, removing configurations amounts to omitting certain singlet patterns and thus breaks the lattice symmetry, which induces doping with visons and ultimately closes the vison gap.  However, as we have observed, the dominating correlation (and thus gap) at the Heisenberg point is given by spinon correlations.  Thus, a suitable ansatz would have to drive the system into criticality through doping with spinons.
To this end, we would have to resort to a different approach and allow for longer-range singlets e.g.\ by introducing teleportation bonds in the PEPS~\cite{wang2013constructing}, which break the $\mathbb Z_2$ symmetry. We leave the study of such an ansatz for future work.

\vspace*{3.5ex}

\begin{acknowledgments}
We thank Ji-Yao Chen, Henrik Dreyer, and Frank Pollmann for valuable discussions,
and C\'ecile Repellin for providing us with the DMRG data of Ref.~\cite{repellin2017stability}.  
MI and NS acknowledge support by the European Union's Horizon 2020 programme through the ERC Starting Grant WASCOSYS (Grant No.~636201) and from the DFG (German Research Foundation) under Germany's Excellence Strategy (EXC-2111 -- 390814868). Computations have been carried out on the TQO cluster of the Max-Planck-Institute of Quantum Optics.
DP acknowledges support by the TNSTRONG  ANR-16-CE30-0025  and TNTOP ANR-18-CE30-0026-01 grants awarded by the French  Research  Council. 
This work was granted access to the HPC resources of CALMIP supercomputing center under the allocations P1231. 
\end{acknowledgments}

\onecolumngrid
\quad
\twocolumngrid
\quad

\clearpage

\appendix
\section{Energy densities, convergence, extrapolation}\label{app:energy-densities}

First, we report in this appendix the energy densities of optimal states within the simplex RVB ansatz for the breathing hamiltonian. Then we discuss the convergence of energies with an increasing bond dimension of environment tensors. In the end, we discuss a possible extrapolation of energies for $p\to\infty$.

\begin {table}[b!]
\begin{center}
	\bgroup
	\def\arraystretch{1.4}%
	\begin{tabular}{c || l | l | l} 
		$\gamma$  & \quad $p = $ 1 & \quad $p = $ 2 & \quad $p = $ 3 \\ 
		\hline\hline
		0.02 & \ -0.252498053(7) \  &  \ -0.2525466(2) 	\      &  \ -0.2526672(7) \ \\ \hline 
		0.04 & \ -0.25502425(1)  \  &  \ -0.2551382(3)  \      &  \ -0.2553798(6) \ \\ \hline 
		0.10 & \ -0.26277062(4)  \  &  \ -0.2631931(9) 	\        & \ -0.263828(1) \ \\ \hline 
		0.20 & \ -0.27622938(7)  \  & \ -0.277585(2) 	\        & \ -0.278971(3) \ \\ \hline 
		0.40 & \ -0.3050715(1)   \  & \  -0.30996(5) 	\        & \ -0.312873(6) \ \\ \hline 
		0.50 & \ -0.3203627(2)   \  &  \ -0.327776(7)  	\        & \ -0.331305(4) \ \\ \hline 
		0.60 & \ -0.3361675(2)   \  &  \ -0.346514(9)  	\        & \ -0.35054(2) \ \\ \hline
		0.70 & \ -0.3524367(2)	 \  &  \ -0.36605(1) 	\		 & \ -0.37045(2) \ \\ \hline 
		0.96 & \ -0.3965909(2)   \  &  \ -0.41978(1)   	\        & \ -0.42472(3) \ \\ \hline 
		1.00 & \ -0.4035897(2)   \  & \ -0.42833(1)   	\        & \  -0.43333(3) \
	\end{tabular}
	\egroup
\end{center}
\caption {Energy densities for different $p$'s and $\gamma$'s. Mean energies have been computed by taking data points with $\chi\geq60$. 
}
\label{tab:ens}
\end {table}

\begin{figure}[!t]
	\centering
	\includegraphics[width=26.5em]{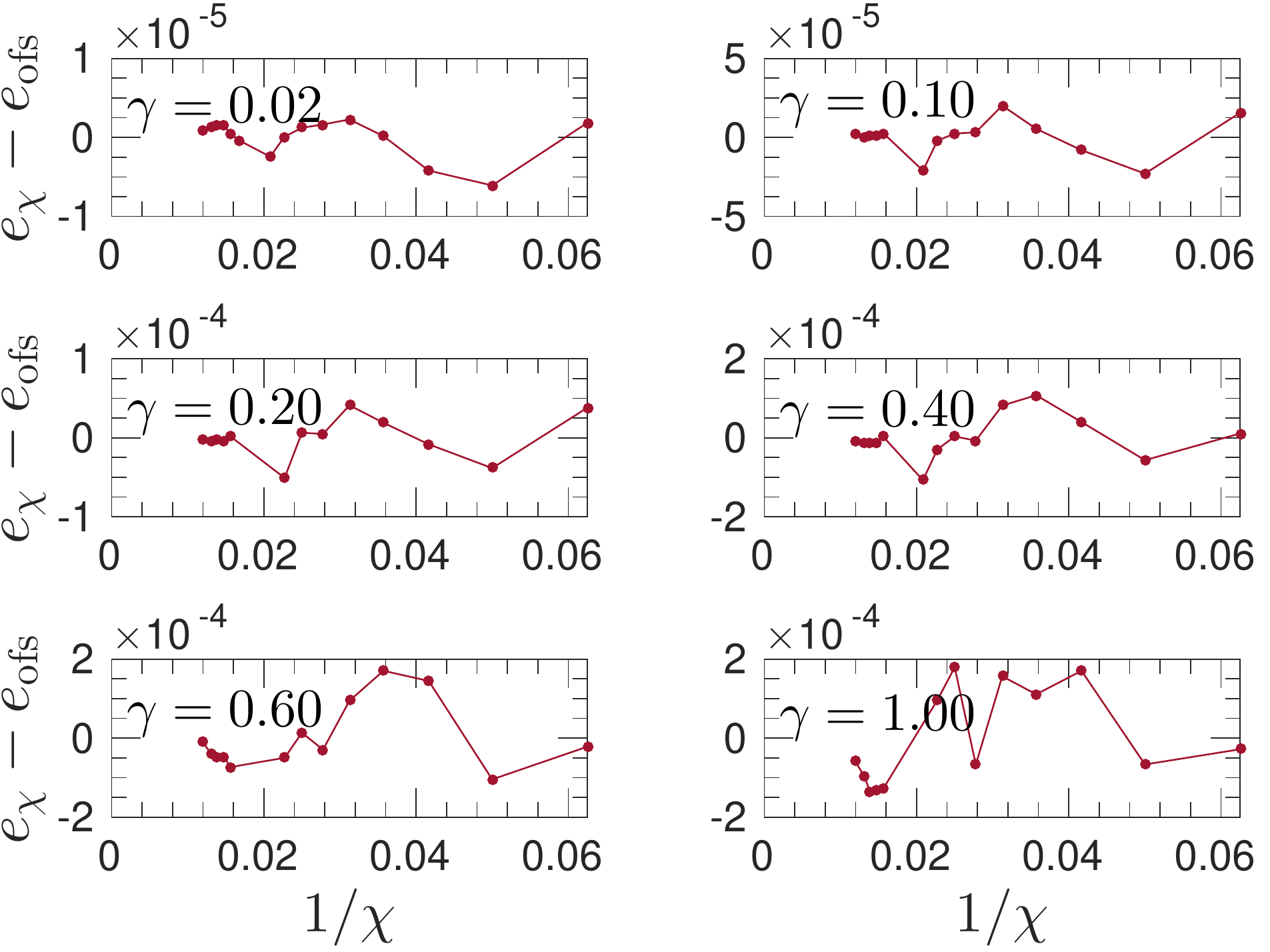}
	\caption{  
	Energy density $e_{\chi}$ as the function of bond dimension $\chi$ for the optimized $p=3$ simplex RVB for different $\gamma$. The offset $e_\mathrm{ofs}$ is chosen by averaging $e_\chi$ for $\chi\geq 16$.  
 We use $e_\chi$ for $\chi \geq 60$ to obtain the mean energies and standard deviations given in Table \ref{tab:ens}, cf.~text.} 
	\label{fig:scal_en}
\end{figure}

We have computed the energy densities for different $\gamma$ and $p$ for $\chi\le 84$. Fig.~\ref{fig:scal_en} shows the behavior of the energy densities $e_\chi$ vs.\ $1/\chi$, relative to an arbitrary offset. We see that the fluctuations decrease by roughly an order of magnitude as we increase $\chi$ to $\chi\ge60$.
We therefore choose to estimate the energy and its error by taking the mean and standard deviation of $e_\chi$ for $\chi\ge60$.
The resulting energy densities are listed in Table \ref{tab:ens}.
We find that energies tend to converge more quickly in the strongly anisotropic regime as compared to near the Heisenberg point; moreover, their convergence is faster for smaller values of $p$. 
Let us note that due to the finite $\chi$, we observe a small splitting in the different bond energies within the left- or right-pointing triangles, of the same magnitude as corresponding flucutations of $e_\chi$ in 
Fig.~\ref{fig:scal_en}.
It is instructive to note that the computation of the expectation values of local observables is more expensive in comparison to the calculation of correlation lengths. The calculation of correlation lengths does not require the whole environment, but only the two fixed points of the transfer operator, which enables us to compute them for larger $\chi$'s~\cite{haegeman2017diagonalizing, iqbal2018study}. 

\begin{figure}[b]
	\centering
	\includegraphics[width=26.5em]{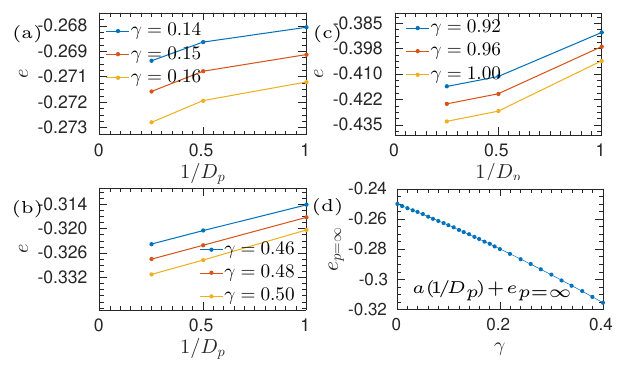}
	\caption{ (a-c) Energies vs.\ inverse bond dimension $1/D_p$ of local tensors in the presence of strong, intermediate, and weak anisotropies. (d) Extrapolated energy densities for small $\gamma$ obtained from a linear fit in $1/D_p$ for $p=1,3$.
	}
	\label{fig:en_extrap_linear}
\end{figure}

As can also be seen from the data in Table  \ref{tab:ens}, the convergence of energies in $p$ is qualitatively different for small and large values of $\gamma$ (Fig. \ref{fig:en_extrap_linear}(a-c)).  The energies exhibit negative (positive) curvature in the presence of strong (weak) anisotropy. One could try to argue that the different scaling behavior of energy densities is a manifestation of the system undergoing a transition from a gapped \z{2} to a gapless SL phase as $\gamma$ increases, but a meaningful assessment would clearly require further data points in $p$. 

Given that we only have $p=1,2,3$, and there might well exist even-odd effects at smaller $p$, extrapolating the energy for $p\to\infty$ is fairly speculative.  One attempt would be based on assuming a gapped phase at small $\gamma$, which suggests an exponential convergence of the energy in the number of Trotter steps (as states above the gap are exponentially supressed) and thus in $p$; this amounts to the common extrapolation in the inverse bond dimension $1/D_p\propto 1/2^p$.
Fig.~\ref{fig:en_extrap_linear}d shows 
$e_{p=\infty}$ obtained from fitting the odd $p$ with $1/2^p$;
the obtained values for small $\gamma$ are in good agreement with the extrapolated DMRG data (see Table~\ref{table:c1}).

\section{Extraction of linear coefficients $c_1$'s}\label{app:c1-coefficient}

\begin{figure}[b]
	\centering
		\includegraphics[width=26.5em]{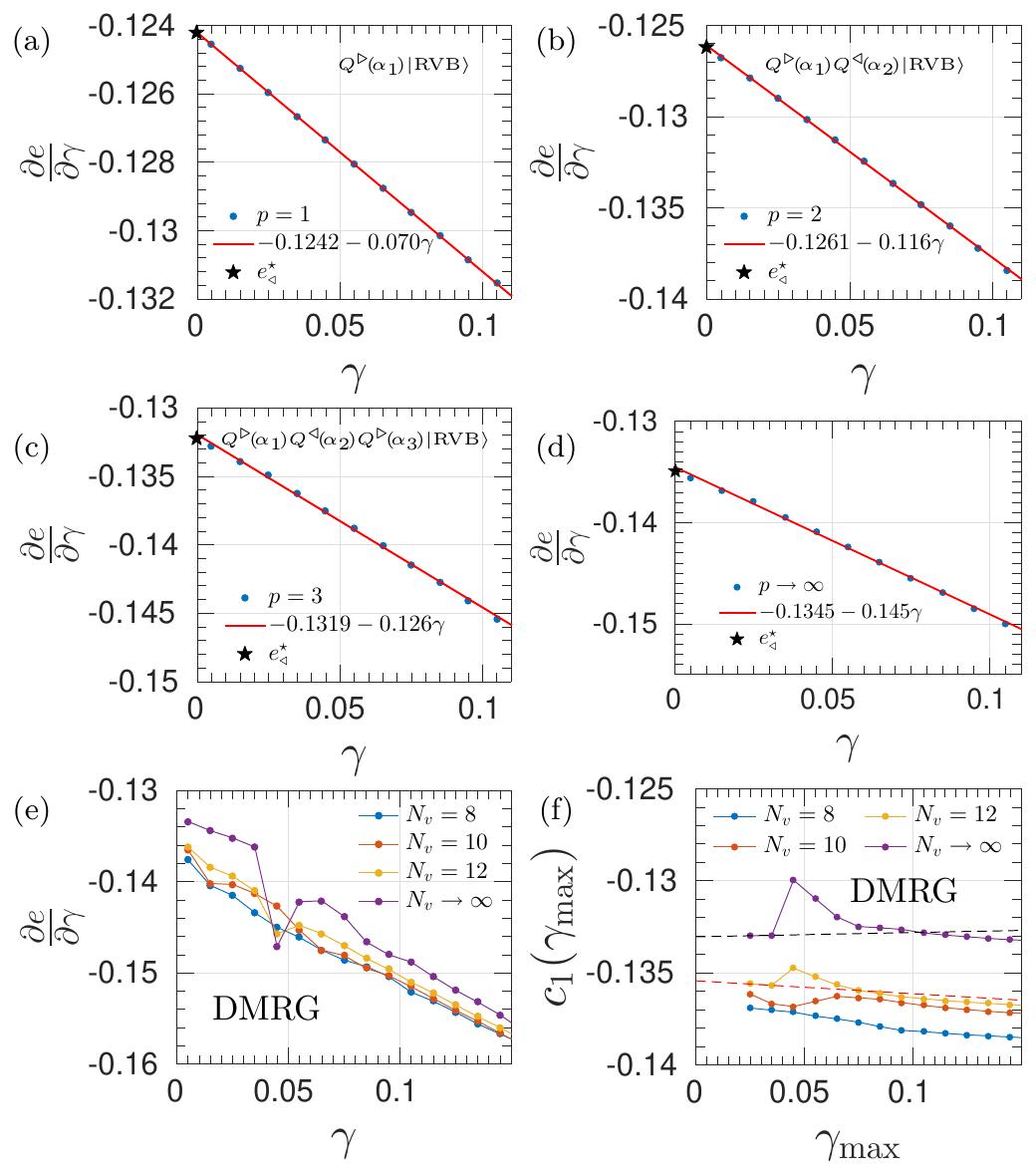}
	\caption{ 
	(a--d) Numerical derivative of the energy density with respect to $\gamma$ in the strongly anisotropic regime for the simplex RVB for $p=1,2,3,\infty$; in (d), the limit $p\to\infty$ has been taken before the derivative. (e)~Derivative of the energy density and (f) coefficient $c_1(\gamma_\mathrm{max})$ for the DMRG data from Ref.~\onlinecite{repellin2017stability}~\cite{cecile:private}; the limit $N\to\infty$ has been taken before the derivative.
	} 
	\label{fig:c1s_gradient}
\end{figure}

In this appendix, we discuss the extraction of the coefficient $c_1$ of the energy expansion $ e(\gamma)=-0.25 + c_1\gamma+ c_2\gamma^2 +\dots \,$ in the strongly anisotropic limit. Since
$c_1 = \lim_{\gamma\to0} \frac{\partial e}{\partial \gamma}, $ given a sufficient number of points, we can reliably take the numerical derivative of the energy density and examine its behavior in the limit $\gamma\to0$. Fig.~\ref{fig:c1s_gradient}(a-d) shows the derivative of energy density as a function of $\gamma$ for the simplex RVB. 
From a linear fit of the data, we can extract $c_1$ for different values of $p$. As an alternative approach, we can also directly fit the energy densities quadratically over the interval $[0,\gamma_{\text{max}}]$, and extract $c_1$ in the limit $\gamma_{\text{max}}\rightarrow 0$. The values of $c_1$ which we get for the simplex RVB by using either the derivative or quadratic fitting are in very close agreement with each other. 

\begin{figure}
	\centering
	\includegraphics[width=21.0em]{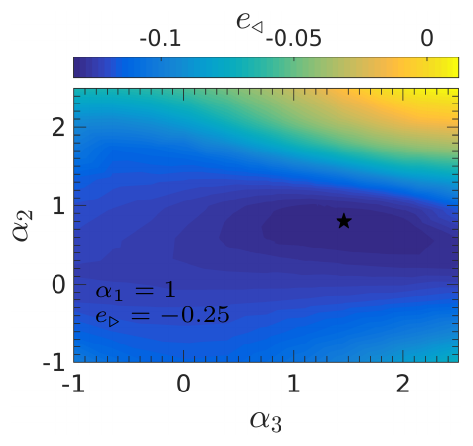}
	\caption{ 
		Energy density of left-pointing triangles as the function of parameters $\alpha_2$ and $\alpha_3$ with the constraint $\alpha_1=1$ for the $p=3$ simpex RVB, used to perturbatively extract $c_1$. Note that the energy landscape possesses no local minima, rendering the optimization stable.
		} 
	\label{fig:energy_surface}
\end{figure}

Furthermore, as explained in Sec.~\ref{sec:results}, we can use perturbation theory to extract $c_1$  for the simplex RVB by fixing $\alpha_1=1$ and minimizing the energy density $e_{\triangleleft}^{\star}\equiv c_1$ of left-pointing triangles. The corresponding $c_1$ are indicated by stars in Fig.~\ref{fig:c1s_gradient}; 
they match well with the values we get by fitting the derivative of energies.
Fig.~\ref{fig:energy_surface} shows the energy density of left-pointing triangles for the simplex RVB with $p=3$ as a function of the remaining two parameters; noteworthily, we find that the energy landscape has no local minima and thus allows for a reliable minimization. Let us emphasize here that in the case of simplex RVB, the perturbative approach for evaluating $c_1$ is more efficient as it eliminates one variable from the variational optimization.

For comparsion, we have extracted the coefficient $c_1$ for the DMRG data from Ref.~\onlinecite{repellin2017stability}~\cite{cecile:private}. Fig.~\ref{fig:c1s_gradient}(e) shows the derivative of the energy density for different \mbox{YC$N_v$-$2$} cylinders, as well as the derivative of the $N_v\to\infty$ energy obtained by 
linear extrapolation in $1/N_v$ (changing the order of limit and derivative gives almost extactly the same values).
However, one needs to be very careful about the extrapolation as the DMRG data suggests a phase transition to the nematic phase in the strongly anisotropic limit, and the phase boundary is sensitive to $N_v$. To extract bounds on $c_1$ from the DMRG data, we therefore examine the behavior of $c_1(\gamma_{\mathrm{max}})$ which is obtained by linear fitting of $\frac{\partial e}{\partial \gamma}$ over the interval $\left[0,\gamma_{\mathrm{max}}\right]$ for both the $N_v=12$ and extrapolated $N_v\rightarrow\infty$ data (Fig.~\ref{fig:c1s_gradient}f). 
A further linear extrapolation of $c_1(\gamma_{\mathrm{max}})$, where we restrict to points $\gamma_\mathrm{max}$ in the nematic phase, gives estimates of $c_1$ for the DMRG data.

\section{Simplex ansatz and vison condensed gapless spin liquid states}\label{app:gaplessrvb}

In RVB-type wavefunctions, $\mathrm U(1)$ symmetries are typically related to critical behavior.  This is most prominent in the square lattice dimer or RVB model, which can be mapped to a height model and thus a critical $\mathrm U(1)$ field theory \cite{fradkin:dimer-height,fradkin:book-fieldtheory}. Corresponding behavior has also been observed in PEPS, where continuous virtual symmetries (corresponding to the counting of singlets) have been related to critical behavior, both in the context of RVB-type models and beyond \cite{wang:rvb-square-lattice,dreyer2018projected,Gauthe2019}.

In this appendix, we construct a modification of the kagome NN RVB which acquires a U(1) symmetry, and show that it yields a gapless spin liquid  state; subsequently, we use this state to build a candidate for a critical simplex RVB and analyze its suitability as a candidate for the breathing KHAFM. 
 For the PEPS description, we begin by considering the blocked local tensor of the NN RVB wavefunction. It is defined on a three-site unit cell containingone right-pointing triangle and is of the form
 \begin{equation}\label{eq:gaplessRVB}
\begin{split}
A(\lambda) &= \left(\cbox{15.0}{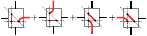} \right) + \\
& \ \ \lambda\left(\cbox{15.0}{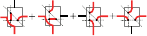} \right)
\end{split}
\end{equation}
with $\lambda=1$. 
Here, each term in the sum can be regarded as a tile, marked with a ``bond'' color (red or black) at each edge. The tiling rule is that only edges with equal color can be connected. If we subsequently ignore the black edges, this results in a pattern where pairs of neighboring vertices on the kagome lattice are connected by red lines: If we replace those by singlets, this precisely yields a NN singlet covering, and it is straightforward to check that all NN coverings areobtained this way. (At a technical level, the bond dimension is $D=3$, where black legs correspond to a $\ket2$ ``no singlet'' state, while red legs are in the states $\ket0$, $\ket1$, and 
perfectly correlated with the corresponding physical spin.  Pairs of red vertices within a unit cell are placed in a singlet, and contraction involves an $i\sigma_y$ in the $\{\ket0,\ket1\}$ subspace, 
see Refs.~\onlinecite{verstraete:comp-power-of-peps,schuch2012resonating}).

The red legs clearly obey a $\mathbb Z_2$ Gauss law with a $-1$ charge per unit cell.  We can now introduce a generalization of the NN RVB by changing $\lambda$, which corresponds to changing the relative weight of different singlet coverings.  In particular, if we choose $\lambda\equiv0$, we are left only with terms which all have exactly one red leg: The tensor $A(\lambda\equiv0)$ has acquired a (staggered) $\mathrm U(1)$ Gauss law, in analogy to the square lattice RVB state, suggestive of a critical behavior at $\gamma=0$.

\begin{figure}[t]
	\centering
	\includegraphics[width=26.5em]{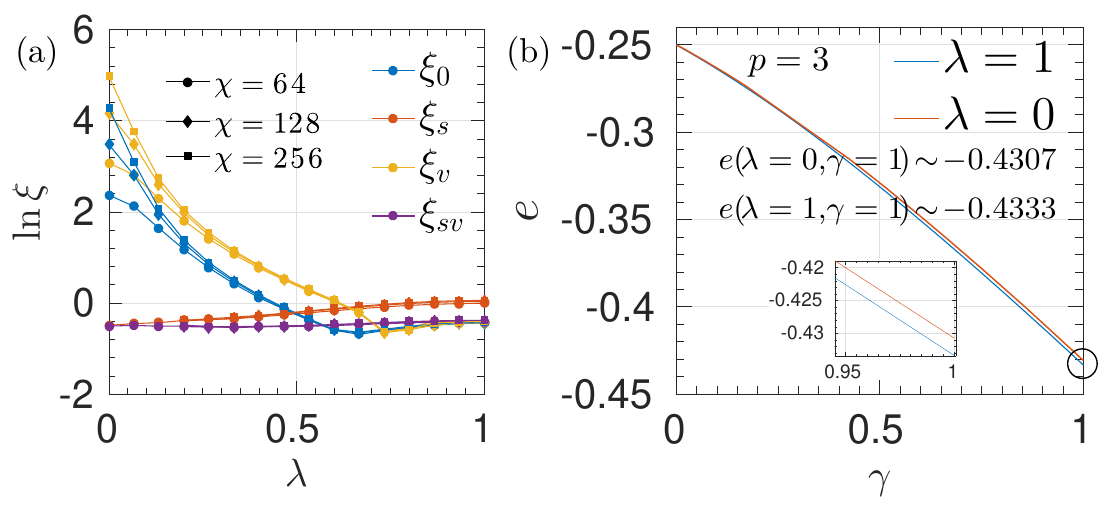}
	\caption{ 
	(a) Correlation lengths of topologically trivial and non-trivial excitations for the $\mathrm{U}(1)$-invariant variant of the NN RVB state as defined in Eq.~\eqref{eq:gaplessRVB} with $\lambda=0$. 
	(b) Optimal energy density for the $p=3$ simplex RVB built on top of the RVB with $A(\lambda=0)$ and $A(\lambda=1)$. } 
	\label{fig:U1}
\end{figure}

We start by analyzing the behavior of the ``bare'' RVB as we change $\lambda$.  Fig.~\ref{fig:U1}a shows the correlation lengths for trivial and anyon-anyon correlations for the different anyon sectors. We find that as we approach $\lambda\to0$, the vison and trivial correlations diverge; this is consistent with the interpretation that changing $\lambda$, which introduces a disbalance between different singlet configurations, dopes the systems with visons which ultimately makes the vison gap close and gives rise to criticality
 due to vison condensation. The critical nature at $\lambda=0$ is confirmed by analyzing the dependence of the correlation length at $\lambda=0$ on $\chi$, which exhibits a rapid growth. Note that we observe criticality only at $\lambda=0$, which is quite different from the model in Ref.~\cite{JiYaoChen2018} where an extended critical region around the $\mathrm U(1)$ point is reported.

We now use the $\mathrm U(1)$ RVB (i.e., at $\lambda=0$) to construct a simplex RVB, that is, we follow Eq.~\eqref{eq:simplex-ansatz}, but with the $\lambda=0$ RVB as a starting state. We have computed the optimal energies for $p=3$, shown in Fig.~\ref{fig:U1}b. However, we find that the energies are above the energies for the simplex RVB ansatz built on top of the NN RVB state, in particular in the vicinity of the Heisenberg point. This is consistent with the fact that we expect a closing of the spinon gap to be the driving mechanism behind a potential critical spin liquid at the Heisenberg point, while our ansatz at $\lambda=0$ exhibits criticality to due a closing vison gap. For the same reason, the application of $Q$'s tends to increase the spinon correlations at the cost of decreasing vison correlations, thus effectively driving the ansatz away from criticality (at least initially); this can also be understood from the fact that the $Q$'s permute singlets and thus can restore the singlet patterns which are missing in the RVB at $\lambda=0$.


\begin{thebibliography}{52}%
\makeatletter
\providecommand \@ifxundefined [1]{%
 \@ifx{#1\undefined}
}%
\providecommand \@ifnum [1]{%
 \ifnum #1\expandafter \@firstoftwo
 \else \expandafter \@secondoftwo
 \fi
}%
\providecommand \@ifx [1]{%
 \ifx #1\expandafter \@firstoftwo
 \else \expandafter \@secondoftwo
 \fi
}%
\providecommand \natexlab [1]{#1}%
\providecommand \enquote  [1]{``#1''}%
\providecommand \bibnamefont  [1]{#1}%
\providecommand \bibfnamefont [1]{#1}%
\providecommand \citenamefont [1]{#1}%
\providecommand \href@noop [0]{\@secondoftwo}%
\providecommand \href [0]{\begingroup \@sanitize@url \@href}%
\providecommand \@href[1]{\@@startlink{#1}\@@href}%
\providecommand \@@href[1]{\endgroup#1\@@endlink}%
\providecommand \@sanitize@url [0]{\catcode `\\12\catcode `\$12\catcode
  `\&12\catcode `\#12\catcode `\^12\catcode `\_12\catcode `\%12\relax}%
\providecommand \@@startlink[1]{}%
\providecommand \@@endlink[0]{}%
\providecommand \url  [0]{\begingroup\@sanitize@url \@url }%
\providecommand \@url [1]{\endgroup\@href {#1}{\urlprefix }}%
\providecommand \urlprefix  [0]{URL }%
\providecommand \Eprint [0]{\href }%
\providecommand \doibase [0]{http://dx.doi.org/}%
\providecommand \selectlanguage [0]{\@gobble}%
\providecommand \bibinfo  [0]{\@secondoftwo}%
\providecommand \bibfield  [0]{\@secondoftwo}%
\providecommand \translation [1]{[#1]}%
\providecommand \BibitemOpen [0]{}%
\providecommand \bibitemStop [0]{}%
\providecommand \bibitemNoStop [0]{.\EOS\space}%
\providecommand \EOS [0]{\spacefactor3000\relax}%
\providecommand \BibitemShut  [1]{\csname bibitem#1\endcsname}%
\let\auto@bib@innerbib\@empty
\bibitem [{\citenamefont {Anderson}(1973)}]{Anderson1973}%
  \BibitemOpen
  \bibfield  {author} {\bibinfo {author} {\bibfnamefont {P.~W.}\ \bibnamefont
  {Anderson}},\ }\href {\doibase
  http://dx.doi.org/10.1016/0025-5408(73)90167-0} {\bibfield  {journal}
  {\bibinfo  {journal} {Materials Research Bulletin}\ }\textbf {\bibinfo
  {volume} {8}},\ \bibinfo {pages} {153 } (\bibinfo {year} {1973})}\BibitemShut
  {NoStop}%
\bibitem [{\citenamefont {Anderson}(1987)}]{RVB}%
  \BibitemOpen
  \bibfield  {author} {\bibinfo {author} {\bibfnamefont {P.~W.}\ \bibnamefont
  {Anderson}},\ }\href {\doibase 10.1126/science.235.4793.1196} {\bibfield
  {journal} {\bibinfo  {journal} {Science}\ }\textbf {\bibinfo {volume}
  {235}},\ \bibinfo {pages} {1196} (\bibinfo {year} {1987})}\BibitemShut
  {NoStop}%
\bibitem [{\citenamefont {Poilblanc}\ \emph {et~al.}(2012)\citenamefont
  {Poilblanc}, \citenamefont {Schuch}, \citenamefont {P\'erez-Garc\'{\i}a},\
  and\ \citenamefont {Cirac}}]{poilblanc2012topological}%
  \BibitemOpen
  \bibfield  {author} {\bibinfo {author} {\bibfnamefont {D.}~\bibnamefont
  {Poilblanc}}, \bibinfo {author} {\bibfnamefont {N.}~\bibnamefont {Schuch}},
  \bibinfo {author} {\bibfnamefont {D.}~\bibnamefont {P\'erez-Garc\'{\i}a}}, \
  and\ \bibinfo {author} {\bibfnamefont {J.~I.}\ \bibnamefont {Cirac}},\ }\href
  {\doibase 10.1103/PhysRevB.86.014404} {\bibfield  {journal} {\bibinfo
  {journal} {Phys. Rev. B}\ }\textbf {\bibinfo {volume} {86}},\ \bibinfo
  {pages} {014404} (\bibinfo {year} {2012})}\BibitemShut {NoStop}%
\bibitem [{\citenamefont {Schuch}\ \emph {et~al.}(2012)\citenamefont {Schuch},
  \citenamefont {Poilblanc}, \citenamefont {Cirac},\ and\ \citenamefont
  {P\'erez-Garc\'{\i}a}}]{schuch2012resonating}%
  \BibitemOpen
  \bibfield  {author} {\bibinfo {author} {\bibfnamefont {N.}~\bibnamefont
  {Schuch}}, \bibinfo {author} {\bibfnamefont {D.}~\bibnamefont {Poilblanc}},
  \bibinfo {author} {\bibfnamefont {J.~I.}\ \bibnamefont {Cirac}}, \ and\
  \bibinfo {author} {\bibfnamefont {D.}~\bibnamefont {P\'erez-Garc\'{\i}a}},\
  }\href {\doibase 10.1103/PhysRevB.86.115108} {\bibfield  {journal} {\bibinfo
  {journal} {Phys. Rev. B}\ }\textbf {\bibinfo {volume} {86}},\ \bibinfo
  {pages} {115108} (\bibinfo {year} {2012})}\BibitemShut {NoStop}%
\bibitem [{\citenamefont {Wildeboer}\ and\ \citenamefont
  {Seidel}(2012)}]{wildeboer}%
  \BibitemOpen
  \bibfield  {author} {\bibinfo {author} {\bibfnamefont {J.}~\bibnamefont
  {Wildeboer}}\ and\ \bibinfo {author} {\bibfnamefont {A.}~\bibnamefont
  {Seidel}},\ }\href {\doibase 10.1103/PhysRevLett.109.147208} {\bibfield
  {journal} {\bibinfo  {journal} {Phys. Rev. Lett.}\ }\textbf {\bibinfo
  {volume} {109}},\ \bibinfo {pages} {147208} (\bibinfo {year}
  {2012})}\BibitemShut {NoStop}%
\bibitem [{\citenamefont {Yang}\ and\ \citenamefont {Yao}(2012)}]{yangfan}%
  \BibitemOpen
  \bibfield  {author} {\bibinfo {author} {\bibfnamefont {F.}~\bibnamefont
  {Yang}}\ and\ \bibinfo {author} {\bibfnamefont {H.}~\bibnamefont {Yao}},\
  }\href {\doibase 10.1103/PhysRevLett.109.147209} {\bibfield  {journal}
  {\bibinfo  {journal} {Phys. Rev. Lett.}\ }\textbf {\bibinfo {volume} {109}},\
  \bibinfo {pages} {147209} (\bibinfo {year} {2012})}\BibitemShut {NoStop}%
\bibitem [{\citenamefont {Zhou}\ \emph {et~al.}(2014)\citenamefont {Zhou},
  \citenamefont {Wildeboer},\ and\ \citenamefont {Seidel}}]{zhou2014}%
  \BibitemOpen
  \bibfield  {author} {\bibinfo {author} {\bibfnamefont {Z.}~\bibnamefont
  {Zhou}}, \bibinfo {author} {\bibfnamefont {J.}~\bibnamefont {Wildeboer}}, \
  and\ \bibinfo {author} {\bibfnamefont {A.}~\bibnamefont {Seidel}},\ }\href
  {\doibase 10.1103/PhysRevB.89.035123} {\bibfield  {journal} {\bibinfo
  {journal} {Phys. Rev. B}\ }\textbf {\bibinfo {volume} {89}},\ \bibinfo
  {pages} {035123} (\bibinfo {year} {2014})},\ \Eprint
  {http://arxiv.org/abs/arXiv:1310.8000} {arXiv:1310.8000} \BibitemShut
  {NoStop}%
\bibitem [{\citenamefont {Hastings}(2004)}]{hastings2004lieb}%
  \BibitemOpen
  \bibfield  {author} {\bibinfo {author} {\bibfnamefont {M.~B.}\ \bibnamefont
  {Hastings}},\ }\href {\doibase 10.1103/PhysRevB.69.104431} {\bibfield
  {journal} {\bibinfo  {journal} {Phys. Rev. B}\ }\textbf {\bibinfo {volume}
  {69}},\ \bibinfo {pages} {104431} (\bibinfo {year} {2004})}\BibitemShut
  {NoStop}%
\bibitem [{\citenamefont {Ran}\ \emph {et~al.}(2007)\citenamefont {Ran},
  \citenamefont {Hermele}, \citenamefont {Lee},\ and\ \citenamefont
  {Wen}}]{Ran2007}%
  \BibitemOpen
  \bibfield  {author} {\bibinfo {author} {\bibfnamefont {Y.}~\bibnamefont
  {Ran}}, \bibinfo {author} {\bibfnamefont {M.}~\bibnamefont {Hermele}},
  \bibinfo {author} {\bibfnamefont {P.~A.}\ \bibnamefont {Lee}}, \ and\
  \bibinfo {author} {\bibfnamefont {X.-G.}\ \bibnamefont {Wen}},\ }\href
  {\doibase 10.1103/PhysRevLett.98.117205} {\bibfield  {journal} {\bibinfo
  {journal} {Phys. Rev. Lett.}\ }\textbf {\bibinfo {volume} {98}},\ \bibinfo
  {pages} {117205} (\bibinfo {year} {2007})}\BibitemShut {NoStop}%
\bibitem [{\citenamefont {Iqbal}\ \emph {et~al.}(2013)\citenamefont {Iqbal},
  \citenamefont {Becca}, \citenamefont {Sorella},\ and\ \citenamefont
  {Poilblanc}}]{iqbal2013gapless}%
  \BibitemOpen
  \bibfield  {author} {\bibinfo {author} {\bibfnamefont {Y.}~\bibnamefont
  {Iqbal}}, \bibinfo {author} {\bibfnamefont {F.}~\bibnamefont {Becca}},
  \bibinfo {author} {\bibfnamefont {S.}~\bibnamefont {Sorella}}, \ and\
  \bibinfo {author} {\bibfnamefont {D.}~\bibnamefont {Poilblanc}},\ }\href
  {\doibase 10.1103/PhysRevB.87.060405} {\bibfield  {journal} {\bibinfo
  {journal} {Phys. Rev. B}\ }\textbf {\bibinfo {volume} {87}},\ \bibinfo
  {pages} {060405} (\bibinfo {year} {2013})}\BibitemShut {NoStop}%
\bibitem [{\citenamefont {He}\ \emph {et~al.}(2017)\citenamefont {He},
  \citenamefont {Zaletel}, \citenamefont {Oshikawa},\ and\ \citenamefont
  {Pollmann}}]{he2017signatures}%
  \BibitemOpen
  \bibfield  {author} {\bibinfo {author} {\bibfnamefont {Y.-C.}\ \bibnamefont
  {He}}, \bibinfo {author} {\bibfnamefont {M.~P.}\ \bibnamefont {Zaletel}},
  \bibinfo {author} {\bibfnamefont {M.}~\bibnamefont {Oshikawa}}, \ and\
  \bibinfo {author} {\bibfnamefont {F.}~\bibnamefont {Pollmann}},\ }\href
  {\doibase 10.1103/PhysRevX.7.031020} {\bibfield  {journal} {\bibinfo
  {journal} {Phys. Rev. X}\ }\textbf {\bibinfo {volume} {7}},\ \bibinfo {pages}
  {031020} (\bibinfo {year} {2017})}\BibitemShut {NoStop}%
\bibitem [{\citenamefont {Liao}\ \emph {et~al.}(2017)\citenamefont {Liao},
  \citenamefont {Xie}, \citenamefont {Chen}, \citenamefont {Liu}, \citenamefont
  {Xie}, \citenamefont {Huang}, \citenamefont {Normand},\ and\ \citenamefont
  {Xiang}}]{liao2017gapless}%
  \BibitemOpen
  \bibfield  {author} {\bibinfo {author} {\bibfnamefont {H.~J.}\ \bibnamefont
  {Liao}}, \bibinfo {author} {\bibfnamefont {Z.~Y.}\ \bibnamefont {Xie}},
  \bibinfo {author} {\bibfnamefont {J.}~\bibnamefont {Chen}}, \bibinfo {author}
  {\bibfnamefont {Z.~Y.}\ \bibnamefont {Liu}}, \bibinfo {author} {\bibfnamefont
  {H.~D.}\ \bibnamefont {Xie}}, \bibinfo {author} {\bibfnamefont {R.~Z.}\
  \bibnamefont {Huang}}, \bibinfo {author} {\bibfnamefont {B.}~\bibnamefont
  {Normand}}, \ and\ \bibinfo {author} {\bibfnamefont {T.}~\bibnamefont
  {Xiang}},\ }\href {\doibase 10.1103/PhysRevLett.118.137202} {\bibfield
  {journal} {\bibinfo  {journal} {Phys. Rev. Lett.}\ }\textbf {\bibinfo
  {volume} {118}},\ \bibinfo {pages} {137202} (\bibinfo {year}
  {2017})}\BibitemShut {NoStop}%
\bibitem [{\citenamefont {Yan}\ \emph {et~al.}(2011)\citenamefont {Yan},
  \citenamefont {Huse},\ and\ \citenamefont {White}}]{yan2011spin}%
  \BibitemOpen
  \bibfield  {author} {\bibinfo {author} {\bibfnamefont {S.}~\bibnamefont
  {Yan}}, \bibinfo {author} {\bibfnamefont {D.~A.}\ \bibnamefont {Huse}}, \
  and\ \bibinfo {author} {\bibfnamefont {S.~R.}\ \bibnamefont {White}},\ }\href
  {\doibase 10.1126/science.1201080} {\bibfield  {journal} {\bibinfo  {journal}
  {Science}\ }\textbf {\bibinfo {volume} {332}},\ \bibinfo {pages} {1173}
  (\bibinfo {year} {2011})}\BibitemShut {NoStop}%
\bibitem [{\citenamefont {Mei}\ \emph {et~al.}(2017)\citenamefont {Mei},
  \citenamefont {Chen}, \citenamefont {He},\ and\ \citenamefont
  {Wen}}]{mei2017gapped}%
  \BibitemOpen
  \bibfield  {author} {\bibinfo {author} {\bibfnamefont {J.-W.}\ \bibnamefont
  {Mei}}, \bibinfo {author} {\bibfnamefont {J.-Y.}\ \bibnamefont {Chen}},
  \bibinfo {author} {\bibfnamefont {H.}~\bibnamefont {He}}, \ and\ \bibinfo
  {author} {\bibfnamefont {X.-G.}\ \bibnamefont {Wen}},\ }\href {\doibase
  10.1103/PhysRevB.95.235107} {\bibfield  {journal} {\bibinfo  {journal} {Phys.
  Rev. B}\ }\textbf {\bibinfo {volume} {95}},\ \bibinfo {pages} {235107}
  (\bibinfo {year} {2017})}\BibitemShut {NoStop}%
\bibitem [{\citenamefont {Depenbrock}\ \emph {et~al.}(2012)\citenamefont
  {Depenbrock}, \citenamefont {McCulloch},\ and\ \citenamefont
  {Schollwoeck}}]{depenbrock:kagome-heisenberg-dmrg}%
  \BibitemOpen
  \bibfield  {author} {\bibinfo {author} {\bibfnamefont {S.}~\bibnamefont
  {Depenbrock}}, \bibinfo {author} {\bibfnamefont {I.~P.}\ \bibnamefont
  {McCulloch}}, \ and\ \bibinfo {author} {\bibfnamefont {U.}~\bibnamefont
  {Schollwoeck}},\ }\href@noop {} {\bibfield  {journal} {\bibinfo  {journal}
  {Phys. Rev. Lett.}\ }\textbf {\bibinfo {volume} {109}},\ \bibinfo {pages}
  {067201} (\bibinfo {year} {2012})},\ \Eprint
  {http://arxiv.org/abs/arXiv:1205.4858} {arXiv:1205.4858} \BibitemShut
  {NoStop}%
\bibitem [{\citenamefont {Okamoto}\ \emph {et~al.}(2013)\citenamefont
  {Okamoto}, \citenamefont {Nilsen}, \citenamefont {Attfield},\ and\
  \citenamefont {Hiroi}}]{okamoto2013breathing}%
  \BibitemOpen
  \bibfield  {author} {\bibinfo {author} {\bibfnamefont {Y.}~\bibnamefont
  {Okamoto}}, \bibinfo {author} {\bibfnamefont {G.~J.}\ \bibnamefont {Nilsen}},
  \bibinfo {author} {\bibfnamefont {J.~P.}\ \bibnamefont {Attfield}}, \ and\
  \bibinfo {author} {\bibfnamefont {Z.}~\bibnamefont {Hiroi}},\ }\href
  {\doibase 10.1103/PhysRevLett.110.097203} {\bibfield  {journal} {\bibinfo
  {journal} {Phys. Rev. Lett.}\ }\textbf {\bibinfo {volume} {110}},\ \bibinfo
  {pages} {097203} (\bibinfo {year} {2013})}\BibitemShut {NoStop}%
\bibitem [{\citenamefont {Aidoudi}\ \emph {et~al.}(2011)\citenamefont
  {Aidoudi}, \citenamefont {Aldous}, \citenamefont {Goff}, \citenamefont
  {Slawin}, \citenamefont {Attfield}, \citenamefont {Morris},\ and\
  \citenamefont {Lightfoot}}]{Aidoudi2011}%
  \BibitemOpen
  \bibfield  {author} {\bibinfo {author} {\bibfnamefont {F.~H.}\ \bibnamefont
  {Aidoudi}}, \bibinfo {author} {\bibfnamefont {D.~W.}\ \bibnamefont {Aldous}},
  \bibinfo {author} {\bibfnamefont {R.~J.}\ \bibnamefont {Goff}}, \bibinfo
  {author} {\bibfnamefont {A.~M.~Z.}\ \bibnamefont {Slawin}}, \bibinfo {author}
  {\bibfnamefont {J.~P.}\ \bibnamefont {Attfield}}, \bibinfo {author}
  {\bibfnamefont {R.~E.}\ \bibnamefont {Morris}}, \ and\ \bibinfo {author}
  {\bibfnamefont {P.}~\bibnamefont {Lightfoot}},\ }\href {\doibase
  10.1038/nchem.1129} {\bibfield  {journal} {\bibinfo  {journal} {Nature
  Chemistry}\ }\textbf {\bibinfo {volume} {3}},\ \bibinfo {pages} {801}
  (\bibinfo {year} {2011})}\BibitemShut {NoStop}%
\bibitem [{\citenamefont {Clark}\ \emph {et~al.}(2013)\citenamefont {Clark},
  \citenamefont {Orain}, \citenamefont {Bert}, \citenamefont {De~Vries},
  \citenamefont {Aidoudi}, \citenamefont {Morris}, \citenamefont {Lightfoot},
  \citenamefont {Lord}, \citenamefont {Telling}, \citenamefont {Bonville},
  \citenamefont {Attfield}, \citenamefont {Mendels},\ and\ \citenamefont
  {Harrison}}]{Harrison2013}%
  \BibitemOpen
  \bibfield  {author} {\bibinfo {author} {\bibfnamefont {L.}~\bibnamefont
  {Clark}}, \bibinfo {author} {\bibfnamefont {J.~C.}\ \bibnamefont {Orain}},
  \bibinfo {author} {\bibfnamefont {F.}~\bibnamefont {Bert}}, \bibinfo {author}
  {\bibfnamefont {M.~A.}\ \bibnamefont {De~Vries}}, \bibinfo {author}
  {\bibfnamefont {F.~H.}\ \bibnamefont {Aidoudi}}, \bibinfo {author}
  {\bibfnamefont {R.~E.}\ \bibnamefont {Morris}}, \bibinfo {author}
  {\bibfnamefont {P.}~\bibnamefont {Lightfoot}}, \bibinfo {author}
  {\bibfnamefont {J.~S.}\ \bibnamefont {Lord}}, \bibinfo {author}
  {\bibfnamefont {M.~T.~F.}\ \bibnamefont {Telling}}, \bibinfo {author}
  {\bibfnamefont {P.}~\bibnamefont {Bonville}}, \bibinfo {author}
  {\bibfnamefont {J.~P.}\ \bibnamefont {Attfield}}, \bibinfo {author}
  {\bibfnamefont {P.}~\bibnamefont {Mendels}}, \ and\ \bibinfo {author}
  {\bibfnamefont {A.}~\bibnamefont {Harrison}},\ }\href {\doibase
  10.1103/PhysRevLett.110.207208} {\bibfield  {journal} {\bibinfo  {journal}
  {Phys. Rev. Lett.}\ }\textbf {\bibinfo {volume} {110}},\ \bibinfo {pages}
  {207208} (\bibinfo {year} {2013})}\BibitemShut {NoStop}%
\bibitem [{\citenamefont {Orain}\ \emph {et~al.}(2017)\citenamefont {Orain},
  \citenamefont {Bernu}, \citenamefont {Mendels}, \citenamefont {Clark},
  \citenamefont {Aidoudi}, \citenamefont {Lightfoot}, \citenamefont {Morris},\
  and\ \citenamefont {Bert}}]{Bert2017}%
  \BibitemOpen
  \bibfield  {author} {\bibinfo {author} {\bibfnamefont {J.-C.}\ \bibnamefont
  {Orain}}, \bibinfo {author} {\bibfnamefont {B.}~\bibnamefont {Bernu}},
  \bibinfo {author} {\bibfnamefont {P.}~\bibnamefont {Mendels}}, \bibinfo
  {author} {\bibfnamefont {L.}~\bibnamefont {Clark}}, \bibinfo {author}
  {\bibfnamefont {F.~H.}\ \bibnamefont {Aidoudi}}, \bibinfo {author}
  {\bibfnamefont {P.}~\bibnamefont {Lightfoot}}, \bibinfo {author}
  {\bibfnamefont {R.~E.}\ \bibnamefont {Morris}}, \ and\ \bibinfo {author}
  {\bibfnamefont {F.}~\bibnamefont {Bert}},\ }\href {\doibase
  10.1103/PhysRevLett.118.237203} {\bibfield  {journal} {\bibinfo  {journal}
  {Phys. Rev. Lett.}\ }\textbf {\bibinfo {volume} {118}},\ \bibinfo {pages}
  {237203} (\bibinfo {year} {2017})}\BibitemShut {NoStop}%
\bibitem [{\citenamefont {Mila}(1998)}]{mila:breathing}%
  \BibitemOpen
  \bibfield  {author} {\bibinfo {author} {\bibfnamefont {F.}~\bibnamefont
  {Mila}},\ }\href@noop {} {\bibfield  {journal} {\bibinfo  {journal} {Phys.
  Rev. Lett.}\ }\textbf {\bibinfo {volume} {81}},\ \bibinfo {pages} {2356}
  (\bibinfo {year} {1998})},\ \Eprint {http://arxiv.org/abs/cond-mat/9805078}
  {cond-mat/9805078} \BibitemShut {NoStop}%
\bibitem [{\citenamefont {Schaffer}\ \emph {et~al.}(2017)\citenamefont
  {Schaffer}, \citenamefont {Huh}, \citenamefont {Hwang},\ and\ \citenamefont
  {Kim}}]{schaffer2017quantum}%
  \BibitemOpen
  \bibfield  {author} {\bibinfo {author} {\bibfnamefont {R.}~\bibnamefont
  {Schaffer}}, \bibinfo {author} {\bibfnamefont {Y.}~\bibnamefont {Huh}},
  \bibinfo {author} {\bibfnamefont {K.}~\bibnamefont {Hwang}}, \ and\ \bibinfo
  {author} {\bibfnamefont {Y.~B.}\ \bibnamefont {Kim}},\ }\href {\doibase
  10.1103/PhysRevB.95.054410} {\bibfield  {journal} {\bibinfo  {journal} {Phys.
  Rev. B}\ }\textbf {\bibinfo {volume} {95}},\ \bibinfo {pages} {054410}
  (\bibinfo {year} {2017})}\BibitemShut {NoStop}%
\bibitem [{\citenamefont {Iqbal}\ \emph
  {et~al.}(2018{\natexlab{a}})\citenamefont {Iqbal}, \citenamefont {Poilblanc},
  \citenamefont {Thomale},\ and\ \citenamefont {Becca}}]{iqbal2018persistence}%
  \BibitemOpen
  \bibfield  {author} {\bibinfo {author} {\bibfnamefont {Y.}~\bibnamefont
  {Iqbal}}, \bibinfo {author} {\bibfnamefont {D.}~\bibnamefont {Poilblanc}},
  \bibinfo {author} {\bibfnamefont {R.}~\bibnamefont {Thomale}}, \ and\
  \bibinfo {author} {\bibfnamefont {F.}~\bibnamefont {Becca}},\ }\href
  {\doibase 10.1103/PhysRevB.97.115127} {\bibfield  {journal} {\bibinfo
  {journal} {Phys. Rev. B}\ }\textbf {\bibinfo {volume} {97}},\ \bibinfo
  {pages} {115127} (\bibinfo {year} {2018}{\natexlab{a}})}\BibitemShut
  {NoStop}%
\bibitem [{\citenamefont {Repellin}\ \emph {et~al.}(2017)\citenamefont
  {Repellin}, \citenamefont {He},\ and\ \citenamefont
  {Pollmann}}]{repellin2017stability}%
  \BibitemOpen
  \bibfield  {author} {\bibinfo {author} {\bibfnamefont {C.}~\bibnamefont
  {Repellin}}, \bibinfo {author} {\bibfnamefont {Y.-C.}\ \bibnamefont {He}}, \
  and\ \bibinfo {author} {\bibfnamefont {F.}~\bibnamefont {Pollmann}},\ }\href
  {\doibase 10.1103/PhysRevB.96.205124} {\bibfield  {journal} {\bibinfo
  {journal} {Phys. Rev. B}\ }\textbf {\bibinfo {volume} {96}},\ \bibinfo
  {pages} {205124} (\bibinfo {year} {2017})}\BibitemShut {NoStop}%
\bibitem [{\citenamefont {Vanderstraeten}\ \emph {et~al.}(2017)\citenamefont
  {Vanderstraeten}, \citenamefont {Marien}, \citenamefont {Haegeman},
  \citenamefont {Schuch}, \citenamefont {Vidal},\ and\ \citenamefont
  {Verstraete}}]{vanderstraeten:peps-perturbations}%
  \BibitemOpen
  \bibfield  {author} {\bibinfo {author} {\bibfnamefont {L.}~\bibnamefont
  {Vanderstraeten}}, \bibinfo {author} {\bibfnamefont {M.}~\bibnamefont
  {Marien}}, \bibinfo {author} {\bibfnamefont {J.}~\bibnamefont {Haegeman}},
  \bibinfo {author} {\bibfnamefont {N.}~\bibnamefont {Schuch}}, \bibinfo
  {author} {\bibfnamefont {J.}~\bibnamefont {Vidal}}, \ and\ \bibinfo {author}
  {\bibfnamefont {F.}~\bibnamefont {Verstraete}},\ }\href {\doibase
  10.1103/physrevlett.119.070401} {\bibfield  {journal} {\bibinfo  {journal}
  {Physical Review Letters}\ }\textbf {\bibinfo {volume} {119}} (\bibinfo
  {year} {2017}),\ 10.1103/physrevlett.119.070401},\ \Eprint
  {http://arxiv.org/abs/1703.04112} {1703.04112} \BibitemShut {NoStop}%
\bibitem [{\citenamefont {Verstraete}\ and\ \citenamefont
  {Cirac}(2004)}]{verstraete2004renormalization}%
  \BibitemOpen
  \bibfield  {author} {\bibinfo {author} {\bibfnamefont {F.}~\bibnamefont
  {Verstraete}}\ and\ \bibinfo {author} {\bibfnamefont {J.~I.}\ \bibnamefont
  {Cirac}},\ }\href {https://arxiv.org/abs/cond-mat/0407066} {\bibfield
  {journal} {\bibinfo  {journal} {arXiv preprint cond-mat/0407066}\ } (\bibinfo
  {year} {2004})}\BibitemShut {NoStop}%
\bibitem [{\citenamefont {Verstraete}\ \emph {et~al.}(2006)\citenamefont
  {Verstraete}, \citenamefont {Wolf}, \citenamefont {Perez-Garcia},\ and\
  \citenamefont {Cirac}}]{verstraete:comp-power-of-peps}%
  \BibitemOpen
  \bibfield  {author} {\bibinfo {author} {\bibfnamefont {F.}~\bibnamefont
  {Verstraete}}, \bibinfo {author} {\bibfnamefont {M.~M.}\ \bibnamefont
  {Wolf}}, \bibinfo {author} {\bibfnamefont {D.}~\bibnamefont {Perez-Garcia}},
  \ and\ \bibinfo {author} {\bibfnamefont {J.~I.}\ \bibnamefont {Cirac}},\
  }\href@noop {} {\bibfield  {journal} {\bibinfo  {journal} {Phys.\ Rev.\
  Lett.}\ }\textbf {\bibinfo {volume} {96}},\ \bibinfo {pages} {220601}
  (\bibinfo {year} {2006})},\ \Eprint {http://arxiv.org/abs/quant-ph/0601075}
  {quant-ph/0601075} \BibitemShut {NoStop}%
\bibitem [{\citenamefont {Poilblanc}\ and\ \citenamefont
  {Schuch}(2013)}]{poilblanc2013simplex}%
  \BibitemOpen
  \bibfield  {author} {\bibinfo {author} {\bibfnamefont {D.}~\bibnamefont
  {Poilblanc}}\ and\ \bibinfo {author} {\bibfnamefont {N.}~\bibnamefont
  {Schuch}},\ }\href {\doibase 10.1103/PhysRevB.87.140407} {\bibfield
  {journal} {\bibinfo  {journal} {Phys. Rev. B}\ }\textbf {\bibinfo {volume}
  {87}},\ \bibinfo {pages} {140407} (\bibinfo {year} {2013})}\BibitemShut
  {NoStop}%
\bibitem [{\citenamefont {Elser}\ and\ \citenamefont
  {Zeng}(1993)}]{elser1993kagome}%
  \BibitemOpen
  \bibfield  {author} {\bibinfo {author} {\bibfnamefont {V.}~\bibnamefont
  {Elser}}\ and\ \bibinfo {author} {\bibfnamefont {C.}~\bibnamefont {Zeng}},\
  }\href {\doibase 10.1103/PhysRevB.48.13647} {\bibfield  {journal} {\bibinfo
  {journal} {Phys. Rev. B}\ }\textbf {\bibinfo {volume} {48}},\ \bibinfo
  {pages} {13647} (\bibinfo {year} {1993})}\BibitemShut {NoStop}%
\bibitem [{\citenamefont {Zeng}\ and\ \citenamefont
  {Elser}(1995)}]{zeng1995quantum}%
  \BibitemOpen
  \bibfield  {author} {\bibinfo {author} {\bibfnamefont {C.}~\bibnamefont
  {Zeng}}\ and\ \bibinfo {author} {\bibfnamefont {V.}~\bibnamefont {Elser}},\
  }\href {\doibase 10.1103/PhysRevB.51.8318} {\bibfield  {journal} {\bibinfo
  {journal} {Phys. Rev. B}\ }\textbf {\bibinfo {volume} {51}},\ \bibinfo
  {pages} {8318} (\bibinfo {year} {1995})}\BibitemShut {NoStop}%
\bibitem [{\citenamefont {Hastings}(2006)}]{hastings:locally}%
  \BibitemOpen
  \bibfield  {author} {\bibinfo {author} {\bibfnamefont {M.~B.}\ \bibnamefont
  {Hastings}},\ }\href@noop {} {\bibfield  {journal} {\bibinfo  {journal}
  {Phys. Rev. B}\ }\textbf {\bibinfo {volume} {73}},\ \bibinfo {pages} {085115}
  (\bibinfo {year} {2006})},\ \Eprint {http://arxiv.org/abs/cond-mat/0508554}
  {cond-mat/0508554} \BibitemShut {NoStop}%
\bibitem [{\citenamefont {Molnar}\ \emph {et~al.}(2015)\citenamefont {Molnar},
  \citenamefont {Schuch}, \citenamefont {Verstraete},\ and\ \citenamefont
  {Cirac}}]{molnar:thermal-peps}%
  \BibitemOpen
  \bibfield  {author} {\bibinfo {author} {\bibfnamefont {A.}~\bibnamefont
  {Molnar}}, \bibinfo {author} {\bibfnamefont {N.}~\bibnamefont {Schuch}},
  \bibinfo {author} {\bibfnamefont {F.}~\bibnamefont {Verstraete}}, \ and\
  \bibinfo {author} {\bibfnamefont {J.~I.}\ \bibnamefont {Cirac}},\ }\href@noop
  {} {\bibfield  {journal} {\bibinfo  {journal} {Phys. Rev. B}\ }\textbf
  {\bibinfo {volume} {91}},\ \bibinfo {pages} {045138} (\bibinfo {year}
  {2015})},\ \Eprint {http://arxiv.org/abs/arXiv:1406.2973} {arXiv:1406.2973}
  \BibitemShut {NoStop}%
\bibitem [{Note1()}]{Note1}%
  \BibitemOpen
  \bibinfo {note} {The same compression can also be obtained with the opposite
  blocking, noting that in expectation values, the final $Q^\triangleright $
  appears as ${Q^\triangleright (\alpha _i)}^\dagger Q^\triangleright (\alpha
  _i)\propto Q^\triangleright (2\alpha _i-\alpha _i^2)$, which can be
  implemented with bond dimension $4$, i.e.\ $2$ per ket/bra
  layer.}\BibitemShut {Stop}%
\bibitem [{\citenamefont {Jordan}\ \emph {et~al.}(2008)\citenamefont {Jordan},
  \citenamefont {Orus}, \citenamefont {Vidal}, \citenamefont {Verstraete},\
  and\ \citenamefont {Cirac}}]{jordan:iPEPS}%
  \BibitemOpen
  \bibfield  {author} {\bibinfo {author} {\bibfnamefont {J.}~\bibnamefont
  {Jordan}}, \bibinfo {author} {\bibfnamefont {R.}~\bibnamefont {Orus}},
  \bibinfo {author} {\bibfnamefont {G.}~\bibnamefont {Vidal}}, \bibinfo
  {author} {\bibfnamefont {F.}~\bibnamefont {Verstraete}}, \ and\ \bibinfo
  {author} {\bibfnamefont {J.~I.}\ \bibnamefont {Cirac}},\ }\href {\doibase
  10.1103/PhysRevLett.101.250602} {\bibfield  {journal} {\bibinfo  {journal}
  {Phys. Rev. Lett.}\ }\textbf {\bibinfo {volume} {101}},\ \bibinfo {pages}
  {250602} (\bibinfo {year} {2008})},\ \Eprint
  {http://arxiv.org/abs/cond-mat/0703788} {cond-mat/0703788} \BibitemShut
  {NoStop}%
\bibitem [{\citenamefont {Haegeman}\ and\ \citenamefont
  {Verstraete}(2017)}]{haegeman2017diagonalizing}%
  \BibitemOpen
  \bibfield  {author} {\bibinfo {author} {\bibfnamefont {J.}~\bibnamefont
  {Haegeman}}\ and\ \bibinfo {author} {\bibfnamefont {F.}~\bibnamefont
  {Verstraete}},\ }\href
  {https://doi.org/10.1146/annurev-conmatphys-031016-025507} {\bibfield
  {journal} {\bibinfo  {journal} {Annual Review of Condensed Matter Physics}\
  }\textbf {\bibinfo {volume} {8}},\ \bibinfo {pages} {355} (\bibinfo {year}
  {2017})}\BibitemShut {NoStop}%
\bibitem [{\citenamefont {Duivenvoorden}\ \emph {et~al.}(2017)\citenamefont
  {Duivenvoorden}, \citenamefont {Iqbal}, \citenamefont {Haegeman},
  \citenamefont {Verstraete},\ and\ \citenamefont
  {Schuch}}]{duivenvoorden2017entanglement}%
  \BibitemOpen
  \bibfield  {author} {\bibinfo {author} {\bibfnamefont {K.}~\bibnamefont
  {Duivenvoorden}}, \bibinfo {author} {\bibfnamefont {M.}~\bibnamefont
  {Iqbal}}, \bibinfo {author} {\bibfnamefont {J.}~\bibnamefont {Haegeman}},
  \bibinfo {author} {\bibfnamefont {F.}~\bibnamefont {Verstraete}}, \ and\
  \bibinfo {author} {\bibfnamefont {N.}~\bibnamefont {Schuch}},\ }\href
  {\doibase 10.1103/PhysRevB.95.235119} {\bibfield  {journal} {\bibinfo
  {journal} {Phys. Rev. B}\ }\textbf {\bibinfo {volume} {95}},\ \bibinfo
  {pages} {235119} (\bibinfo {year} {2017})}\BibitemShut {NoStop}%
\bibitem [{\citenamefont {Iqbal}\ \emph
  {et~al.}(2018{\natexlab{b}})\citenamefont {Iqbal}, \citenamefont
  {Duivenvoorden},\ and\ \citenamefont {Schuch}}]{iqbal2018study}%
  \BibitemOpen
  \bibfield  {author} {\bibinfo {author} {\bibfnamefont {M.}~\bibnamefont
  {Iqbal}}, \bibinfo {author} {\bibfnamefont {K.}~\bibnamefont
  {Duivenvoorden}}, \ and\ \bibinfo {author} {\bibfnamefont {N.}~\bibnamefont
  {Schuch}},\ }\href {\doibase 10.1103/PhysRevB.97.195124} {\bibfield
  {journal} {\bibinfo  {journal} {Phys. Rev. B}\ }\textbf {\bibinfo {volume}
  {97}},\ \bibinfo {pages} {195124} (\bibinfo {year}
  {2018}{\natexlab{b}})}\BibitemShut {NoStop}%
\bibitem [{\citenamefont {{C.\ Repellin, private
  communication.}}()}]{cecile:private}%
  \BibitemOpen
  \bibinfo {author} {\bibnamefont {{C.\ Repellin, private
  communication.}}}\BibitemShut {Stop}%
\bibitem [{Note2()}]{Note2}%
  \BibitemOpen
\bibfield  {author} {  }\bibinfo {note} {In fact, the two tests probe the same
  property: Equal energies for both triangles are equivalent to zero slope at
  $\delta =0$, since otherwise, the wavefunction for some non-zero $\delta $
  would constitute a better ansatz even at $\delta =0$.}\BibitemShut {Stop}%
\bibitem [{\citenamefont {{Chen}}\ \emph {et~al.}(2010)\citenamefont {{Chen}},
  \citenamefont {{Zeng}}, \citenamefont {{Gu}}, \citenamefont {{Chuang}},\ and\
  \citenamefont {{Wen}}}]{chen:topo-symmetry-conditions}%
  \BibitemOpen
  \bibfield  {author} {\bibinfo {author} {\bibfnamefont {X.}~\bibnamefont
  {{Chen}}}, \bibinfo {author} {\bibfnamefont {B.}~\bibnamefont {{Zeng}}},
  \bibinfo {author} {\bibfnamefont {Z.}~\bibnamefont {{Gu}}}, \bibinfo {author}
  {\bibfnamefont {I.~L.}\ \bibnamefont {{Chuang}}}, \ and\ \bibinfo {author}
  {\bibfnamefont {X.}~\bibnamefont {{Wen}}},\ }\href {\doibase
  10.1103/PhysRevB.82.165119} {\bibfield  {journal} {\bibinfo  {journal} {Phys.
  Rev. B}\ }\textbf {\bibinfo {volume} {82}},\ \bibinfo {pages} {165119}
  (\bibinfo {year} {2010})},\ \Eprint {http://arxiv.org/abs/arXiv:1003.1774}
  {arXiv:1003.1774} \BibitemShut {NoStop}%
\bibitem [{\citenamefont {{Schuch}}\ \emph {et~al.}(2010)\citenamefont
  {{Schuch}}, \citenamefont {{Cirac}},\ and\ \citenamefont
  {{P{\'e}rez-Garc{\'{\i}}a}}}]{schuch:peps-sym}%
  \BibitemOpen
  \bibfield  {author} {\bibinfo {author} {\bibfnamefont {N.}~\bibnamefont
  {{Schuch}}}, \bibinfo {author} {\bibfnamefont {I.}~\bibnamefont {{Cirac}}}, \
  and\ \bibinfo {author} {\bibfnamefont {D.}~\bibnamefont
  {{P{\'e}rez-Garc{\'{\i}}a}}},\ }\href {\doibase 10.1016/j.aop.2010.05.008}
  {\bibfield  {journal} {\bibinfo  {journal} {Ann. Phys.}\ }\textbf {\bibinfo
  {volume} {325}},\ \bibinfo {pages} {2153} (\bibinfo {year} {2010})},\ \Eprint
  {http://arxiv.org/abs/arXiv:1001.3807} {arXiv:1001.3807} \BibitemShut
  {NoStop}%
\bibitem [{\citenamefont {Schuch}\ \emph {et~al.}(2013)\citenamefont {Schuch},
  \citenamefont {Poilblanc}, \citenamefont {Cirac},\ and\ \citenamefont
  {Perez-Garcia}}]{schuch2013topological}%
  \BibitemOpen
  \bibfield  {author} {\bibinfo {author} {\bibfnamefont {N.}~\bibnamefont
  {Schuch}}, \bibinfo {author} {\bibfnamefont {D.}~\bibnamefont {Poilblanc}},
  \bibinfo {author} {\bibfnamefont {J.~I.}\ \bibnamefont {Cirac}}, \ and\
  \bibinfo {author} {\bibfnamefont {D.}~\bibnamefont {Perez-Garcia}},\
  }\href@noop {} {\bibfield  {journal} {\bibinfo  {journal} {Physical review
  letters}\ }\textbf {\bibinfo {volume} {111}},\ \bibinfo {pages} {090501}
  (\bibinfo {year} {2013})}\BibitemShut {NoStop}%
\bibitem [{\citenamefont {Haegeman}\ \emph {et~al.}(2015)\citenamefont
  {Haegeman}, \citenamefont {Zauner}, \citenamefont {Schuch},\ and\
  \citenamefont {Verstraete}}]{haegeman2015shadows}%
  \BibitemOpen
  \bibfield  {author} {\bibinfo {author} {\bibfnamefont {J.}~\bibnamefont
  {Haegeman}}, \bibinfo {author} {\bibfnamefont {V.}~\bibnamefont {Zauner}},
  \bibinfo {author} {\bibfnamefont {N.}~\bibnamefont {Schuch}}, \ and\ \bibinfo
  {author} {\bibfnamefont {F.}~\bibnamefont {Verstraete}},\ }\href
  {https://doi.org/10.1038/ncomms9284} {\bibfield  {journal} {\bibinfo
  {journal} {Nature communications}\ }\textbf {\bibinfo {volume} {6}} (\bibinfo
  {year} {2015})}\BibitemShut {NoStop}%
\bibitem [{\citenamefont {Iqbal}\ \emph {et~al.}()\citenamefont {Iqbal},
  \citenamefont {Casademunt},\ and\ \citenamefont
  {Schuch}}]{iqbal:rvb-perturb}%
  \BibitemOpen
  \bibfield  {author} {\bibinfo {author} {\bibfnamefont {M.}~\bibnamefont
  {Iqbal}}, \bibinfo {author} {\bibfnamefont {H.}~\bibnamefont {Casademunt}}, \
  and\ \bibinfo {author} {\bibfnamefont {N.}~\bibnamefont {Schuch}},\
  }\href@noop {} {\ }\Eprint {http://arxiv.org/abs/arXiv:1910.06355}
  {arXiv:1910.06355} \BibitemShut {NoStop}%
\bibitem [{Note3()}]{Note3}%
  \BibitemOpen
  \bibinfo {note} {It is worth noting that the values we obtain for $\xi _s$
  and $\xi _v$ for the NN RVB wavefunction are in remarkable agreement with
  their earlier estimates in Ref.~\protect \rev@citealp {poilblanc2013simplex}
  which had been extracted from the finite-size scaling of the energy splitting
  for the corresponding ground states.}\BibitemShut {Stop}%
\bibitem [{Note4()}]{Note4}%
  \BibitemOpen
  \bibinfo {note} {Note that this is only a qualitative argument, as it does
  not take into accounts cancellations due to different phases, or the scaling
  of the number of loop patterns with the distance.}\BibitemShut {Stop}%
\bibitem [{\citenamefont {Wang}\ \emph
  {et~al.}(2013{\natexlab{a}})\citenamefont {Wang}, \citenamefont {Poilblanc},
  \citenamefont {Gu}, \citenamefont {Wen},\ and\ \citenamefont
  {Verstraete}}]{wang2013constructing}%
  \BibitemOpen
  \bibfield  {author} {\bibinfo {author} {\bibfnamefont {L.}~\bibnamefont
  {Wang}}, \bibinfo {author} {\bibfnamefont {D.}~\bibnamefont {Poilblanc}},
  \bibinfo {author} {\bibfnamefont {Z.-C.}\ \bibnamefont {Gu}}, \bibinfo
  {author} {\bibfnamefont {X.-G.}\ \bibnamefont {Wen}}, \ and\ \bibinfo
  {author} {\bibfnamefont {F.}~\bibnamefont {Verstraete}},\ }\href {\doibase
  10.1103/PhysRevLett.111.037202} {\bibfield  {journal} {\bibinfo  {journal}
  {Phys. Rev. Lett.}\ }\textbf {\bibinfo {volume} {111}},\ \bibinfo {pages}
  {037202} (\bibinfo {year} {2013}{\natexlab{a}})}\BibitemShut {NoStop}%
\bibitem [{\citenamefont {Fradkin}\ and\ \citenamefont
  {Kivelson}(1990)}]{fradkin:dimer-height}%
  \BibitemOpen
  \bibfield  {author} {\bibinfo {author} {\bibfnamefont {E.}~\bibnamefont
  {Fradkin}}\ and\ \bibinfo {author} {\bibfnamefont {S.}~\bibnamefont
  {Kivelson}},\ }\href@noop {} {\bibfield  {journal} {\bibinfo  {journal} {Mod.
  Phys. Lett. B}\ }\textbf {\bibinfo {volume} {4}},\ \bibinfo {pages} {225}
  (\bibinfo {year} {1990})}\BibitemShut {NoStop}%
\bibitem [{\citenamefont {Fradkin}(2013)}]{fradkin:book-fieldtheory}%
  \BibitemOpen
  \bibfield  {author} {\bibinfo {author} {\bibfnamefont {E.}~\bibnamefont
  {Fradkin}},\ }\href {https://books.google.de/books?id=x7\_6MX4ye\_wC} {\emph
  {\bibinfo {title} {Field Theories of Condensed Matter Physics}}},\ Field
  Theories of Condensed Matter Physics\ (\bibinfo  {publisher} {Cambridge
  University Press},\ \bibinfo {year} {2013})\BibitemShut {NoStop}%
\bibitem [{\citenamefont {Wang}\ \emph
  {et~al.}(2013{\natexlab{b}})\citenamefont {Wang}, \citenamefont {Poilblanc},
  \citenamefont {Gu}, \citenamefont {Wen},\ and\ \citenamefont
  {Verstraete}}]{wang:rvb-square-lattice}%
  \BibitemOpen
  \bibfield  {author} {\bibinfo {author} {\bibfnamefont {L.}~\bibnamefont
  {Wang}}, \bibinfo {author} {\bibfnamefont {D.}~\bibnamefont {Poilblanc}},
  \bibinfo {author} {\bibfnamefont {Z.-C.}\ \bibnamefont {Gu}}, \bibinfo
  {author} {\bibfnamefont {X.-G.}\ \bibnamefont {Wen}}, \ and\ \bibinfo
  {author} {\bibfnamefont {F.}~\bibnamefont {Verstraete}},\ }\href@noop {}
  {\bibfield  {journal} {\bibinfo  {journal} {Phys. Rev. Lett.}\ }\textbf
  {\bibinfo {volume} {111}},\ \bibinfo {pages} {037202} (\bibinfo {year}
  {2013}{\natexlab{b}})},\ \Eprint {http://arxiv.org/abs/arXiv:1301.4492}
  {arXiv:1301.4492} \BibitemShut {NoStop}%
\bibitem [{\citenamefont {Dreyer}\ \emph {et~al.}(2018)\citenamefont {Dreyer},
  \citenamefont {Cirac},\ and\ \citenamefont {Schuch}}]{dreyer2018projected}%
  \BibitemOpen
  \bibfield  {author} {\bibinfo {author} {\bibfnamefont {H.}~\bibnamefont
  {Dreyer}}, \bibinfo {author} {\bibfnamefont {J.~I.}\ \bibnamefont {Cirac}}, \
  and\ \bibinfo {author} {\bibfnamefont {N.}~\bibnamefont {Schuch}},\ }\href
  {\doibase 10.1103/PhysRevB.98.115120} {\bibfield  {journal} {\bibinfo
  {journal} {Phys. Rev. B}\ }\textbf {\bibinfo {volume} {98}},\ \bibinfo
  {pages} {115120} (\bibinfo {year} {2018})}\BibitemShut {NoStop}%
\bibitem [{\citenamefont {Gauth\'e}\ \emph {et~al.}(2019)\citenamefont
  {Gauth\'e}, \citenamefont {Capponi},\ and\ \citenamefont
  {Poilblanc}}]{Gauthe2019}%
  \BibitemOpen
  \bibfield  {author} {\bibinfo {author} {\bibfnamefont {O.}~\bibnamefont
  {Gauth\'e}}, \bibinfo {author} {\bibfnamefont {S.}~\bibnamefont {Capponi}}, \
  and\ \bibinfo {author} {\bibfnamefont {D.}~\bibnamefont {Poilblanc}},\ }\href
  {\doibase 10.1103/PhysRevB.99.241112} {\bibfield  {journal} {\bibinfo
  {journal} {Phys. Rev. B}\ }\textbf {\bibinfo {volume} {99}},\ \bibinfo
  {pages} {241112} (\bibinfo {year} {2019})}\BibitemShut {NoStop}%
\bibitem [{\citenamefont {Chen}\ and\ \citenamefont
  {Poilblanc}(2018)}]{JiYaoChen2018}%
  \BibitemOpen
  \bibfield  {author} {\bibinfo {author} {\bibfnamefont {J.-Y.}\ \bibnamefont
  {Chen}}\ and\ \bibinfo {author} {\bibfnamefont {D.}~\bibnamefont
  {Poilblanc}},\ }\href {\doibase 10.1103/PhysRevB.97.161107} {\bibfield
  {journal} {\bibinfo  {journal} {Phys. Rev. B}\ }\textbf {\bibinfo {volume}
  {97}},\ \bibinfo {pages} {161107} (\bibinfo {year} {2018})}\BibitemShut
  {NoStop}%
\end{thebibliography}
\end{document}